\documentclass[aps,pra,twocolumn,notitlepage,nopacs,amsmath,amstex,amssymb,citeautoscript,longbibliography,floatfix,twocolumn,nofootinbib]{revtex4-2}
\usepackage{enumitem}
\usepackage{natbib}
\usepackage[english]{babel}
\usepackage{letltxmacro}
\usepackage{latexsym}
\LetLtxMacro{\ORIGselectlanguage}{\selectlanguage}
\makeatletter
\DeclareRobustCommand{\selectlanguage}[1]{%
  \@ifundefined{alias@\string#1}
    {\ORIGselectlanguage{#1}}
    {\begingroup\edef\x{\endgroup
       \noexpand\ORIGselectlanguage{\@nameuse{alias@#1}}}\x}%
}
\newcommand{\definelanguagealias}[2]{%
  \@namedef{alias@#1}{#2}%
}
\makeatother
\definelanguagealias{en}{english}
\definelanguagealias{English}{english}

\newcommand{\be}{\begin{equation}}
\newcommand{\ee}{\end{equation}}
\newcommand{\bea}{\begin{eqnarray}}
\newcommand{\eea}{\end{eqnarray}}

\DeclareMathOperator{\sign}{sign}

\newcommand{\im}{{\rm i}}

\usepackage{amsthm}

\usepackage{graphicx}
\usepackage{bm}
\usepackage{physics}
\usepackage[dvipsnames]{xcolor}
\usepackage{algorithm}
\usepackage{algpseudocode}
\usepackage{blindtext}
\usepackage{orcidlink}
\usepackage[normalem]{ulem}
\makeatletter
\newcommand{\printfnsymbol}[1]{%
  \textsuperscript{\@fnsymbol{#1}}%
}
\makeatother
\usepackage{hyperref}
\usepackage{soul}
\hypersetup{
    unicode=false,          % non-Latin characters in AcrobatÕs bookmarks
    pdftoolbar=false,        % show AcrobatÕs toolbar?
    pdfmenubar=true,        % show AcrobatÕs menu?
    pdffitwindow=false,     % window fit to page when opened
    pdfstartview={FitH},    % fits the width of the page to the window
    pdftitle={},    % title
    pdfauthor={Raimel A. Medina, Maksym Serbyn},     % author
    pdfsubject={Analytical properties of transition states in the QAOA optimization landscape},   % subject of the document
    pdfcreator={},   % creator of the document
    pdfproducer={}, % producer of the document
    pdfkeywords={QAOA, transition states}, % list of keywords
    pdfnewwindow=true,      % links in new window
    colorlinks=true,       % false: boxed links; true: colored links
    linkcolor=teal,          % color of internal links (change box color with linkbordercolor)
    citecolor=teal,        % color of links to bibliography
    filecolor=teal,      % color of file links
    urlcolor=teal           % color of external links
}

\begin{document}
\title{A Recursive Lower Bound on the Energy Improvement of the Quantum Approximate Optimization Algorithm}
\author{Raimel A. Medina \orcidlink{0000-0002-5383-2869}}
\affiliation{Institute of Science and Technology Austria (ISTA), Am Campus 1, 3400 Klosterneuburg, Austria}
\author{Maksym Serbyn \orcidlink{0000-0002-2399-5827}}
\affiliation{Institute of Science and Technology Austria (ISTA), Am Campus 1, 3400 Klosterneuburg, Austria}

\date{\today}
\begin{abstract}
The Quantum Approximate Optimization Algorithm (QAOA) uses a quantum computer to implement a variational method with $2p$ layers of alternating unitary operators, optimized by a classical computer to minimize a cost function. While rigorous performance guarantees exist for the QAOA at small depths $p$, the behavior at large depths remains less clear, though simulations suggest exponentially fast convergence for certain problems. In this work, we gain insights into the deep QAOA using an analytic expansion of the cost function around transition states, which are stationary points characterized by a unique direction of negative curvature. Transition states are constructed recursively: from a local minimum of the QAOA with $p$ layers we obtain transition states of the QAOA with $p+1$ layers. We construct an analytic estimate of the negative curvature and the corresponding direction in parameter space at each transition state. Our estimates, which do not require knowledge of the Hessian spectrum, reveal that the curvature's exponential decay with $p$ is a fundamental signature of the QAOA state converging to an eigenstate of the cost Hamiltonian, as evidenced by the corresponding decay of the energy variance. Expansion of the QAOA cost function along the negative direction to the quartic order gives a lower bound of the QAOA cost function improvement. We provide physical intuition behind the analytic expressions for the local curvature and quartic expansion coefficient. Our numerical study confirms the accuracy of our approximations and reveals that the obtained bound and the true value of the QAOA cost function gain have a characteristic exponential decrease with the number of layers $p$, with the bound decreasing more rapidly. Our study establishes an analytical framework for studying the QAOA's convergence mechanism in the deep circuit regime through the relationship between landscape geometry and physical observables.
\end{abstract}
\maketitle
\section{Introduction}
Variational quantum algorithms~\cite{Cerezo_2021, Bharti_2022} have emerged as a promising approach to leveraging the capabilities of noisy intermediate-scale quantum (NISQ) devices~\cite{preskill2018quantum}. Among these algorithms, the Quantum Approximate Optimization Algorithm (QAOA)~\cite{farhi2014quantum} and the Variational Quantum Eigensolver (VQE)~\cite{peruzzo2014vqe}  stand out due to their potential for solving optimization problems and quantum chemistry simulations, respectively. The idea is to use the quantum computer in a feedback loop with a classical computer, where it implements a variational wave function that is measured to compute the value of the so-called cost function. This information is then fed into a classical computer where it is processed and the variational wave function is subsequently updated aiming to find a minimum of the cost function, which provides an (approximate) solution to the computationally hard problem. 

In the QAOA, the state is prepared by a $p$-level circuit specified by $2p$ variational parameters. It was shown that even at the lowest circuit depth $p = 1$, QAOA has non-trivial provable performance guarantees~\cite{farhi2014quantum, farhi2015quantum}. The existence of known analytical performance guarantees makes the QAOA --- in contrast to the VQE --- a reference algorithm to explore quantum speedups on NISQ devices. In particular, the QAOA has been the subject of both analytical studies~\cite{farhi2020wholegraph, boulebnane2021predicting, boulebnane2022solving, brandao2018fixed, wurtz2021maxcut,zhou2024statistical, Basso_2022, Basso_2022_SKmodel, Sureshbabu_2024, yao2020policy, zhu2022adaptive} and practical implementations~\cite{Harrigan_2021, weidenfeller2022scaling, wurtz2024solving, Ebadi_2022} for small values of the circuit depth, $p$. These studies suggest that significant gains can be expected as $p$ increases, particularly when $p \geq \ln N$, with $N$ representing the number of qubits involved. However, the behavior of QAOA in this high-depth limit remains largely unexplored. Heuristic numerical studies indicate that while the optimization landscape of QAOA becomes increasingly complex~\cite{crooks2018performance, zhou2018quantum}, a robust initialization strategy can lead to rapid convergence. Unfortunately, most existing strategies for initialization rely on heuristic approaches~\cite{sack2021quantum, zhou2018quantum, jain2021graph}, lacking a rigorous analytical foundation.

Addressing this gap, the recent work~\cite{Sack2022} by present authors and collaborators introduced a \emph{recursive} QAOA initialization strategy based on the concept of transition states. Assuming convergence of QAOA at depth $p$ to a local minimum, Ref.~\cite{Sack2022} analytically constructed $2p+1$ transition states for QAOA at depth $p+1$. These transition states, characterized by a single negative curvature direction, ensure a reduction in the cost function value when used as initialization points. Drawing upon this analytical foundation, Ref.~\cite{Sack2022} proposed a \textsc{Greedy} strategy for sequential QAOA initializations that systematically improve the cost function value with increasing $p$. 
Such a recursive approach is practical even in the limit of large $p$, providing an analytical basis for the QAOA initialization.
While the \textsc{Greedy} strategy comes with guarantees of improvement, it requires computing and diagonalizing the Hessian of the cost function at each transition state, thus increasing the cost of optimization. 

In this work, we focus on obtaining analytical insights into deep QAOA using transition states. To this end, we construct an analytical estimate of the minimum Hessian eigenvalue and corresponding eigenvector at each transition state. These results simplify the \textsc{Greedy} initialization strategy~\cite{Sack2022} by effectively eliminating the need to construct or estimate the Hessian of the cost function. Furthermore, we provide a physical intuition behind the expression for the minimal Hessian eigenvalue at the transition state and relate it to the energy variance of the state prepared by the QAOA circuit. 

The analytical approximation of the Hessian eigenvalue and eigenvector, enables us to expand the QAOA cost function to the fourth order around the transition state. A similar expansion was formulated in Ref.~\cite{bukov_2019}, where it was performed to the third order and applied to the optimal quantum control problem.  Our expansion results in a non-trivial local energy minimum located in the vicinity of the transition state,  thereby giving a \emph{recursive} lower bound on the cost function improvement achievable through optimization.  We check our approximations and bound using numerical simulations of the QAOA on instances of 3-regular unweighted/weighted \textsc{MaxCut} instances with $N=10$ to 22 vertices, and find the analytic estimates of the Hessian properties to be accurate within a percent. The analytically obtained lower bound on the cost function improvement scales correctly with the number of qubits \(N\). However, our bound decays exponentially with \(p\) at a faster rate compared to the numerically obtained cost function improvement.

Our results suggest that although the immediate vicinity of analytic transition states does not contain the true cost function minimum of QAOA at depth $p+1$, it qualitatively captures the ``flattening'' of the QAOA landscape with $p$. We speculate that our analytic results on the energy expansion around the transition states may be expanded into an analytic performance guarantee for the QAOA performance at large circuit depths.  Specifically, our work establishes a lower bound on the energy gain expressed via the fidelity of the prepared QAOA state and the true ground state of the cost Hamiltonian. Provided that one manages to bound the increase in fidelity, or potentially, the decrease in the energy variance, this may lead to the desired performance guarantee. 

The rest of the paper is structured as follows. In Sec.~\ref{Sec:II}, we review the QAOA and the transition states construction introduced in previous work. In Sec.~\ref{Sec:III} we present and verify our estimates for the minimum negative Hessian eigenvalue and the corresponding eigenvector. Furthermore, we discuss numerical results that show a connection between the minimum Hessian eigenvalue and the energy dispersion of the QAOA state. Next, in Sec.~\ref{Sec:IV} we present a lower bound on the energy improvement between local minima of the QAOA at circuit depths $p$ and $p+1$. We also test the tightness of the presented bound and discuss its wide implications for the performance of the QAOA. Finally, in Sec.~\ref{Sec:V} we discuss our results and potential future extensions of our work. Appendices \ref{App:numerics}-\ref{Sec:App-expand} present detailed proofs of our analytical results, as well as supporting numerical simulations.

\section{QAOA and transition states \label{Sec:II}}
In this section, we start with defining the QAOA algorithm as applied to the \textsc{MaxCut} problem and review the analytic construction of transition states from Ref.~\cite{Sack2022}.

\subsection{QAOA and the \textsc{MaxCut}\label{Sec:QAOAintro}}
The QAOA~\cite{farhi2014quantum} was first introduced as a near-term algorithm for approximately solving classical combinatorial optimization problems. Here, we focus on the particular case of the maximum cut \textsc{MaxCut} problem. \textsc{MaxCut} seeks to partition a given (un)weighted graph $\mathcal{G}$ with $n_{\mathcal{E}}(\mathcal{G})$ edges into two groups such that the number of edges (or the sum of their weights, for weighted problems) that connect vertices from different groups are maximized. Finding the \textsc{MaxCut} for a graph with $N$ vertices is equivalent to finding a ground state for the $N$-qubit classical Hamiltonian 
\begin{equation}\label{Eq:ising}
H_C = \sum_{\langle i,j\rangle \in \mathcal{E}} J_{ij} \sigma^z_i \sigma^z_j,
\end{equation} 
with the sum running over a set of graph edges $\mathcal{E}$ with weights $J_{ij}$ and $\sigma^z_i$ being the Pauli-$z$ matrix acting on the $i$-th qubit. We assume that this problem has a unique ground state, denoted as $\ket{E_0}$ (this state is unique in the proper sector of global $Z_2$ symmetry, which we use to improve the efficiency of the numerical simulations). The full spectrum of $H_C$ consists of all product states ordered according to their energies and will be used in what follows as a complete basis, $\ket{E_0}, \ket{E_1},\ldots, \ket{E_{2^N-1}}$. 

\begin{figure*}[ht]
\centering
\includegraphics[width=\linewidth]{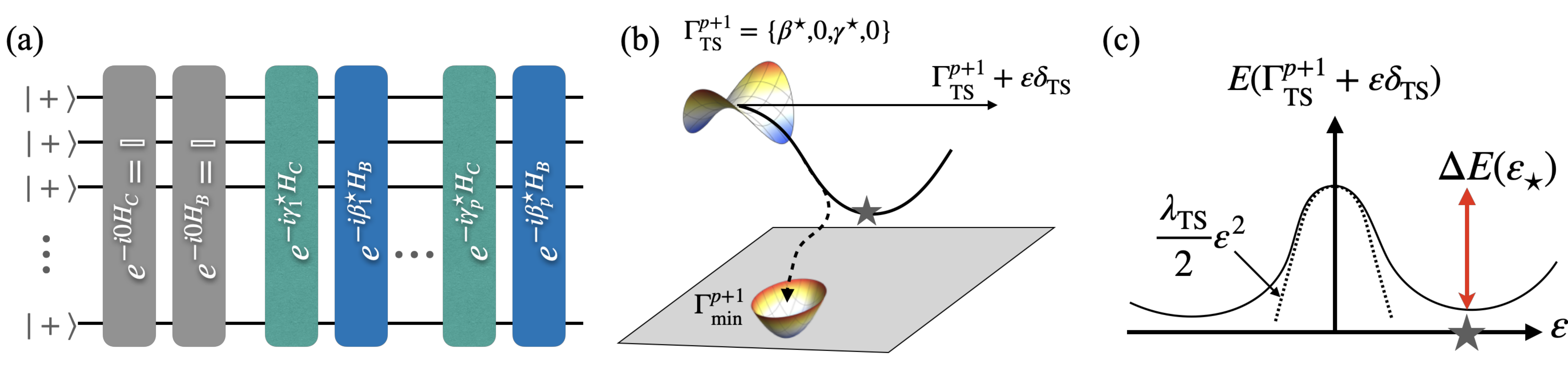}
\caption{(a) Analytic construction of the particular transition state obtained from inserting two identity gates into QAOA$_p$ circuit. (b) We inspect the energy alongside the unique descent direction associated with each of the transition states. The minimum along the unique descent direction (gray star marker) does not correspond to a stationary state of the energy. However, it lower bounds the energy of the minimum obtained by running optimization. (c) Sketch of the projected dependence of the cost function, with $\Delta E(\varepsilon_*)$ putting a rigorous lower bound on the energy improvement at this iteration. 
} 
\label{Fig:cartoon}
\end{figure*}

The depth-$p$ QAOA algorithm~\cite{farhi2014quantum}, denoted in what follows as QAOA$_p$, minimizes the expectation value of the classical Hamiltonian over the variational state $\ket{ \Gamma^p}$ where $\Gamma^p = (\bm{\beta},\bm{\gamma})$ encodes variational angles $\bm{\beta}=(\beta_1,\ldots,\beta_p)$ and $\bm{\gamma}=(\gamma_1,\ldots,\gamma_p)$ shown in  Fig.~\ref{Fig:cartoon}(a):
\begin{equation}\label{eq:QAOA_ansatz}
    \ket{\Gamma^p} = U(\Gamma^p)|+\rangle=\prod_{k=1}^{p} e^{-\im \beta_k H_B} e^{-\im \gamma_k H_C} \ket{+}.
\end{equation}
Here 
\begin{equation}\label{Eq:mixing}
H_B = -\sum_{i=1}^N \sigma^x_i,
\end{equation} 
is the mixing Hamiltonian and the circuit depth $p$ controls the number of classical and mixing Hamiltonian applications. 
The initial product state $\ket{+} =\otimes_{i=1}^N \ket{+}_i$, where all qubits point in the $x$-direction is an equal superposition of all possible graph partitions which is also the ground state of $H_B$. 
Finding the minimum of 
\begin{equation}
E(\Gamma^p)=\bra{ \Gamma^p} H_C \ket{\Gamma^p}
\end{equation}
over angles 
$(\beta_1,\ldots,\beta_p)$ and 
$(\gamma_1,\ldots,\gamma_p)$
that form a set of $2p$ variational parameters,  $\Gamma^p=(\bm{\beta}, \bm{\gamma})$, yields a desired approximation to the ground state of $H_C$, equivalent to an approximate a solution of \textsc{MaxCut}. The scalar function $E(\Gamma^p)$ thus defines a $2p$-dimensional energy landscape where the global minimum yields the best set of QAOA parameters. The performance of the QAOA is typically reported in terms of how close is the approximation ratio to one, 
\begin{equation}
1-r(\Gamma^p)=\frac{E_{0}-E(\Gamma^p)}{E_{0}},
\end{equation}
where $E_{0}$ is the ground state of the classical Hamiltonian~(\ref{Eq:ising}).  From here we see that a decrease in $1-r$ implies that the expectation value of the cost function is approaching the ground state energy of the classical Hamiltonian. 

Here, we restrict our attention to \textsc{MaxCut} on 3-regular graphs, where every vertex is connected to exactly 3 other vertices. In the main text, we focus on unweighted 3-regular graphs, while we delegate the results for weighted 3-regular graphs, with weights $J_{ij}$ chosen uniformly at random from the interval $[0,1]$,  for the Appendix~\ref{Sec:App-ts_1}. It is important to note that the results presented here are fully general (up to algebraic details), meaning that they hold for generic non-commuting Hamiltonians, $H_C$ and $H_B$, provided that the initial state $|\psi_0\rangle$ (or $|+\rangle$ in this work) is an eigenstate of $H_B$.

\subsection{Analytical transition states  \label{Sec:descent}}
Most studies of the QAOA optimization landscape to date were restricted to local minima of the cost function $E(\bm{\beta}, \bm{\gamma})$ since they can be directly obtained using standard gradient-based or some gradient-free optimization routines. Local minima are stationary points of the energy landscape, defined as $\partial_i E(\bm{\beta}, \bm{\gamma})= 0$, with the index $i$ ranging over all $2p$ variational parameters, where all eigenvalues of the Hessian matrix $H_{ij}=\partial_i \partial_j E(\bm{\beta}, \bm{\gamma})$ are positive, that is the Hessian at the local minimum is positive-definite. The other stationary points with $0<k < 2p$ negative Hessian eigenvalues are known as index-$k$ saddle points.  

Work~\cite{Sack2022} analytically constructed index-$1$ saddle points dubbed transition states (TS) hereafter, of the QAOA$_{p+1}$ using a given local minimum of the QAOA$_p$, $\Gamma_{\min}^{p}$. This construction is illustrated in Fig.~\ref{Fig:cartoon}(a), and it consists of the insertion of a pair of identity gates, viewed as additional variational parameters initialized at a value equal to zero. Such insertion is allowed at $2p+1$ possible positions, giving rise to $2p+1$ distinct stationary points of the QAOA$_{p+1}$, $\Gamma_\text{TS}^{p+1}$ with a \emph{unique} negative eigenvalue of the Hessian. We note that while showing that $\Gamma_\text{TS}^{p+1}$ constructed as above are stationary points is relatively straightforward, demonstrating that their Hessian has a single negative eigenvalue is less trivial, with a detailed proof available in Ref.~\cite{Sack2022}.

Given that, by construction, all the $2p+1$ TS have the same energy as the initial local minimum $\Gamma^p_{\min}$, one can use the direction associated with the negative eigenvalue of the Hessian (hereafter referred to as index-1 direction) to further decrease the energy, see Fig.~\ref{Fig:cartoon}(b) for an example. In this way, the TS construction can be used as an initialization scheme that guarantees \emph{improvement} of the QAOA performance with the circuit depth $p$~\cite{Sack2022}. In this work, using the properties of the QAOA energy landscape in the vicinity of the transition states, we quantify the performance improvement of the QAOA by providing a lower bound for the energy improvement after an iteration of the QAOA.

\section{Curvature of energy landscape near transition state\label{Sec:III}}
In this section, we explore the curvature which is the first nontrivial local property of the QAOA energy landscape around the TS constructed out of a local minima $\Gamma^p_{\min}$ of the QAOA$_p$. Using the structure of the Hessian at the TS, we develop an approximation to its unique negative eigenvalue and its corresponding eigenvector. We also uncover connections between the negative curvature of the Hessian at the TS and the excited state population of the prepared QAOA state as a function of the circuit depth $p$.
\subsection{Minimum Hessian eigenvalue and eigenvector}
In this work, our analysis is specifically tailored to the scenario where additional identity gates are incorporated at the initial layer of the pre-existing QAOA$_p$ circuit, as illustrated in Fig.~\ref{Fig:cartoon}(a). This particular scenario corresponds to the transition state denoted as $\Gamma_{\text{TS}}^{p+1}(1,1)$ in Ref.~\cite{Sack2022}. To maintain clarity and avoid unnecessary complexity in notation, we refer to this state more simply throughout our discussion when the context permits. We denote the transition state configuration as:
\begin{equation}
\label{Eq:descent}
\Gamma_{\text{TS}}^{p+1} = (0, \beta_1^\star, \ldots, \beta_p^\star; 0, \gamma_1^\star, \ldots, \gamma_p^\star).
\end{equation}
The primary reason for focusing on this specific transition state is that it significantly simplifies the analytical manipulations required for deriving the worst-case energy improvement achievable through optimizing QAOA$_{p+1}$, starting from a local minimum obtained by QAOA$_{p}$. Additionally, Appendix~\ref{Sec:App-ts_1} presents numerical analyses that benchmark the effectiveness of this approach against the \textsc{Greedy} optimization strategy introduced on~\cite{Sack2022}, which exploits all $2p+1$ transition states derived from $\Gamma^p_{\min}$.

In Appendix~\ref{Sec:App-bound}, we develop a framework for estimating both the minimum Hessian eigenvalue and its corresponding eigenvector. We first establish rigorous lower and upper bounds for the minimum eigenvalue, then extend our analysis to obtain the eigenvector estimate. Our approach leverages a congruence relation between the Hessian at the transition state and a block diagonal matrix, its blocks being: the Hessian at the minimum $H(\Gamma^p_{\min})$ (positive definite) and a $2 \times 2$ block with a negative and a positive eigenvalue. By exploiting this structure, we derive an estimate vector that simultaneously refines our upper bound for the minimum eigenvalue and serves as our eigenvector approximation:
\begin{equation}\label{Eq:negative}
 {\bm \delta}_\text{TS} = \left(-\frac12,\frac12,\underbrace{0,\ldots}_{p-1~\text{zeros}}; -\frac{\sign(b)}{\sqrt2},\underbrace{0,\ldots}_{p~\text{zeros}}\right),
\end{equation}
where the parameter $b$ is the second derivative of the cost function $b=\partial_{\gamma_1}\partial_{\beta_1}E(\Gamma^{p+1}_{\mathrm{TS}})$, which can be expressed as a nested commutator of the following three operators:
\begin{equation}
    \label{eq:b_first_ts}
    b = \langle +| [H_C, [H_B, U^\dagger(\Gamma^p_{\min}) H_C U(\Gamma^p_{\min})]]|+\rangle.
\end{equation}
Notably, these estimates can be computed without prior knowledge of the spectrum of $H(\Gamma^p_{\min})$. While this leads to reduced accuracy in certain cases, as we discuss below, our numerical experiments reveal that even in these scenarios the optimization performance remains unchanged, making these estimates an efficient tool for our subsequent analysis.

The approximate form of the eigenvector (\ref{Eq:negative}) shows that when initialized from the former minima of QAOA$_p$, the classical optimization procedure changes values of angles $\beta_1$, $\gamma_1$ that were initialized at zero initially, as well as the value of $\beta_2=\beta_1^\star$ initialized at the value set by the local minimum at depth $p$. All remaining parameters are left intact at the start of gradient descent. The expression for $b$ above can be further simplified relying on the specific expressions for $H_C$ and $H_B$, using Eq.~\eqref{Eq:ising} and Eq.~\eqref{Eq:mixing} respectively
\begin{equation}
    \label{eq:b_first_ts_final}
    b = 8\langle +| H_C U^\dagger(\Gamma^p_{\min}) H_C U(\Gamma^p_{\min})|+\rangle.
\end{equation}

Finally, we approximate the minimum Hessian eigenvalue $\lambda_{\mathrm{TS}}$ by the expectation value of the Hessian on the approximate eigenvector $\bm{\delta}_{\mathrm{TS}}$, obtaining that it is proportional to the second derivative $b=\partial_{\gamma_1}\partial_{\beta_1}E(\Gamma^{p+1}_{\mathrm{TS}})$ defined above:
\begin{equation}
    \label{eq:curv_first_ts}
    \lambda_{\mathrm{TS}} = - \frac{|b|}{\sqrt{2}} = -4\sqrt{2} |\langle +| H_C U^\dagger(\Gamma^p_{\min}) H_C U(\Gamma^p_{\min})|+\rangle|.
\end{equation}
We refer the discussion of the physical intuition behind this expression to the Sec.~\ref{Sec:cuvature3}, where we show that $\lambda_{\mathrm{TS}}$ vanishes in the case when QAOA unitary circuit rotates the $|+\rangle$ state into an exact eigenstate of $H_C$. It is critical to note that our estimation of the minimum Hessian eigenvalue is potentially computable on a NISQ device. From the form of the transition state specified by Eq.~\eqref{Eq:descent}, we deduce that the value of $b$ in Eq.~\eqref{eq:b_first_ts_final} can be estimated on a NISQ device using additionally $\mathcal{O}(n_{\mathcal{E}}(\mathcal{G}))$ more circuit executions. Here, $n_{\mathcal{E}}(\mathcal{G})$ denotes the number of edges (interaction terms) in the problem graph $\mathcal{G}$ that determines the cost Hamiltonian Eq.~\eqref{Eq:ising}.

Although in this Section we focused on the transition state obtained by padding with zeros the first layer of the QAOA, in the Appendix~\ref{Sec:App-bound}, we show that a similar approach allows us to obtain estimates of eigenvectors and eigenvalues for all the $2p+1$ TS. For a generic transition state $\Gamma^{p+1}_{\mathrm{TS}}(i, j)$, the approximate Hessian eigenvector has non-zero components corresponding to adjacent gates, that is $\beta_i, \beta_{i+1}, \gamma_j$, and  $\gamma_{j-1}$. Moreover, the approximate eigenvalue is given by a particular matrix element of the Hessian when the zeros insertion is at the first or last layer, or by a difference of two particular matrix elements in the Hessian matrix for all remaining transition states. In all the cases, our approximation to the eigenvalue is a measurable quantity, which can be estimated analogously to the procedure described after Eq.~(\ref{eq:curv_first_ts}).

\subsection{Quality of the curvature approximation}
To assess the accuracy of our estimates for the true minimum Hessian eigenvalue and its corresponding eigenvector, we examine a collection of non-isomorphic random instances of 3-regular unweighted graphs with $N=10,\ldots,16$ vertices---18, 34, 55, and 40 instances respectively---executing the QAOA for circuit depths within the range $p \in [1, 30]$. For each local minimum obtained at a given circuit depth $p$, we construct the $2p+1$ transition states and compute their exact numerical Hessians. After determining the Hessian spectrum, we calculate the relative error of our minimum eigenvalue estimate and the deviation of the absolute overlap between the approximate and exact Hessian eigenvector from 1.

\begin{figure}[t]
\centering
         \includegraphics[width=0.99\linewidth]{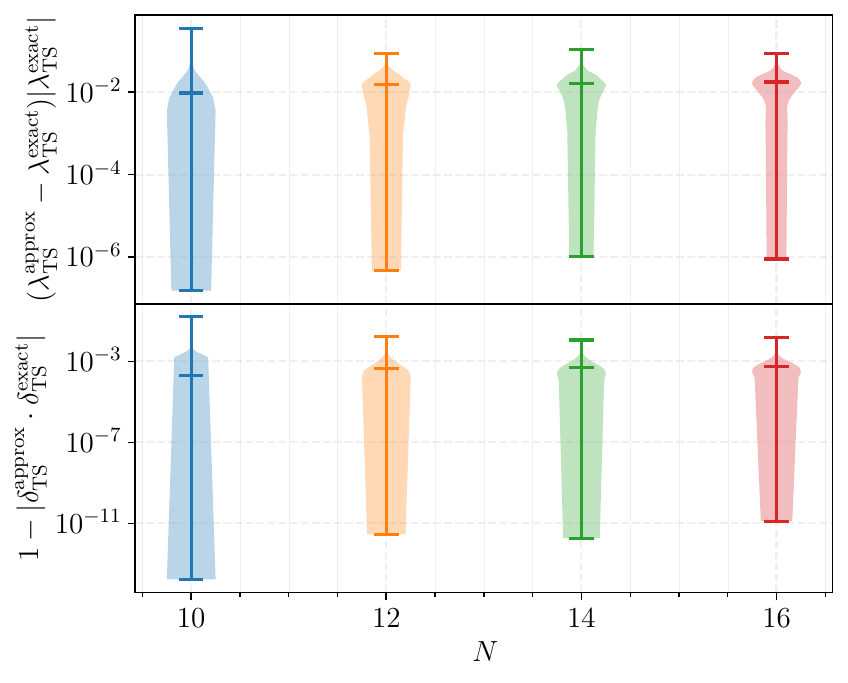}
\caption{Accuracy of curvature and descent direction estimates shown by violin plots for QAOA transition states across graph instances with 10 to 16 vertices and circuit depths ranging from 1 to 30. (\textit{Top}) Relative error in the negative Hessian eigenvalue estimation; the median error is indicated by the horizontal line. (\textit{Bottom}) Deviation from unity in the absolute overlap between the estimated and exact eigenvectors associated with the negative eigenvalue. The shaded regions capture the probability density of the data, reflecting that the accuracy of our eigenvector estimate is consistent across different system sizes.}
\label{Fig:descent}
\end{figure}

We consolidate our findings in Fig.~2, which is composed of ``violin plots'' that illustrate the distribution of the obtained numerical data. The width of each violin indicates the frequency of data points at different error or overlap values, providing insight into the variability of the measures. Typically, the median of the data---indicated by the horizontal line within each violin---reveals that the relative error in the minimum Hessian eigenvalue is on the order of $\mathcal{O}(10^{-2})$, while the deviation from $1$ of the absolute value of the overlap between the exact and approximate eigenstates is on the order of $\mathcal{O}(10^{-3})$. 

The accuracy of our method depends on three key parameters: $b$, $\bar{b}$ and $\kappa_1$. Here, $b$ defined in Eq.~\eqref{eq:b_first_ts} is proportional to the negative Hessian eigenvalue estimate at the transition state $\Gamma^{p+1}_{\mathrm{TS}}(1,1)$, while $\bar{b}$ defined in Eq.~\eqref{eq:bbar_definition} serves the same role for transition states $\Gamma^{p+1}_{\mathrm{TS}}(i,j)$. When either $|b|$ or $|\bar{b}|$ is less than $\kappa_1$ (the smallest eigenvalue of $H(\Gamma^p_{\min})$), the errors remain consistently low ($\sim 1 \%$). For cases where $|b|, |\bar{b}| > \kappa_1$, we observe a decrease in accuracy, however, the thin upper tails of the violin plots indicate that such cases are exceptionally rare, even though they can lead to larger errors (up to $\sim 36\%$ for eigenvalues and $\sim 29\%$ for eigenvectors). Furthermore, in Appendix~\ref{Sec:App-bound} we discuss and verify that such accuracy drops do not impact the optimization performance, with trajectories starting from the same transition state and following either the exact or approximate direction converging to the same local minimum in a comparable number of iterations. This robustness, combined with the visual analysis of the violin plots, underlines the reliability of our method, with the precision of our estimates appearing stable upon increasing the number of qubits $N$.

\subsection{Evolution of the curvature with the depth of QAOA}\label{Sec:cuvature3}
In this section, we focus on the behavior of the curvature with the circuit depth $p$. First, we demonstrate that $\lambda_{\mathrm{TS}}$ is vanishing when QAOA$_p$ prepares an eigenstate of the cost Hamiltonian $H_C$, thereby being proportional to the square root of the \emph{infidelity} of the eigenstate preparation. Moreover, we build a physical intuition for the value of curvature by relating it to the action of the QAOA circuit on the excited states of the mixing Hamiltonian. Next, we suggest the parallel in the behavior of the curvature and the energy variance of the state prepared in the QAOA circuit with respect to the cost Hamiltonian.  Finally, we test our arguments numerically, demonstrating that similarly to the $1-r$, where $r$ is the QAOA approximation ratio, vanishing exponentially with the circuit depth $p$, the curvature and energy variance also decrease exponentially. 

To provide the physical intuition for the value of $\lambda_{\mathrm{TS}}$ in Eq.~(\ref{eq:curv_first_ts}), we write quantum states $U(\Gamma^p_{\min})|+\rangle$ and $U(\Gamma^p_{\min})H_C|+\rangle$ in the following form
\begin{equation}
    \label{eq:expr_for_curv}
    \begin{split}
        U(\Gamma^p_{\min})|+\rangle &= \alpha^0 |E_0 \rangle + \alpha^0_{\perp} |\psi_0\rangle, \\
        U(\Gamma^p_{\min})H_C|+\rangle &= n_C \kappa^0 |E_0 \rangle + n_C\kappa^0_{\perp} |\phi_0\rangle,
    \end{split}
\end{equation}
where we have selected out the ground state of the classical Hamiltonian, $|E_{l=0} \rangle$ (this can be any eigenstate of $H_C$, not necessarily the ground state, as we will discuss later), and remaining states $|\psi_0\rangle$ and $|\phi_0\rangle$ in this expansion are normalized superposition of all other eigenstates of $H_C$, thus being orthogonal to $|E_0 \rangle$ by construction\footnote{The terms in Eq.~\eqref{eq:expr_for_curv} are not completely independent. Using the fact that states $H_C|+\rangle$ and $|+\rangle$ are orthogonal, we can show that following relation holds $\alpha^0 (\kappa^0)^* + \alpha^0_\perp \kappa^0_\perp \langle \phi_0|\psi_0\rangle=0$.}. The constant $n_C=\sum_{\langle ij \rangle }J^2_{ij}$ comes from the norm of $H_C|+\rangle$ and it is added such that $|\kappa^0|^2 + |\kappa^0_{\perp}|^2 = 1$. 

Notations introduced in Eq.~\eqref{eq:expr_for_curv} allow us to rewrite the expression for the curvature, Eq.~\eqref{eq:curv_first_ts}, as
\begin{equation}
\label{eq:curv_refined_expr}
    |\lambda_{\mathrm{TS}}| = 4 \sqrt{2} n_C \alpha^0_{\perp} \kappa^0_{\perp} |\langle \phi_0| H_C - E_0 |\psi_0\rangle|,
\end{equation}
where without loss of generality we assume that factors in this expression, $\alpha^0_{\perp}$ and $ \kappa^0_{\perp}$, are real positive numbers. In notations of Eq.~(\ref{eq:expr_for_curv}) $\alpha^0$ corresponds to the square root of the fidelity of the QAOA prepared state $U(\Gamma^p_{\min})|+\rangle$ to the ground state $|E_0\rangle$ of $H_C$ with eigenvalue $E_0$, and $\alpha^0_\perp = \sqrt{1-|\alpha^0|^2}$ is the infidelity. Crucially, the second line of Eq.~(\ref{eq:expr_for_curv}) defines $\kappa^0$ as square root of the fidelity between states $|E_0\rangle$ and  $U(\Gamma^p_{\min})H_C|+\rangle$, where the latter state physically corresponds to the QAOA circuit applied to an \emph{excited eigenstate} of the mixing Hamiltonian with eigenvalue $-N+4$ (for more general forms of $H_C$ and $H_B$, we expect this state to be a combination of low-lying excited eigenstates of $H_B$). Since the curvature $\lambda_{\mathrm{TS}}$ is proportional to the product of $\alpha^0_{\perp}$ and $\kappa^0_{\perp}$, it implies that it is sensitive not only to the infidelity resulting from the QAOA circuit preparing the desired ground state of the cost Hamiltonian but also to the behavior of the low-lying excited state of the mixing Hamiltonian as it is acted upon by the QAOA circuit.  

From expression~(\ref{eq:curv_refined_expr}), we realize that a non-zero curvature around the transition state $\Gamma^{p+1}_{\mathrm{TS}}$ comes from the fact that the QAOA circuit prepares a superposition of energy eigenstates as quantified by $\alpha^0_\perp \neq 0$ and $\kappa^0_\perp \neq 0$. Furthermore, to determine the curvature we need information on what superposition of eigenstates of the classical Hamiltonian the QAOA unitary creates when acting on the superposition of states $\sigma^z_i \sigma^z_j |+\rangle$, with $\langle i, j \rangle \in \mathcal{E}_G$ corresponding to an edge in the problem graph $G$, that are the superposition of second excited states (two spin flips in $x$ basis) of the mixing Hamiltonian. 

\begin{figure}
    \centering
    \includegraphics[width=0.99\columnwidth]{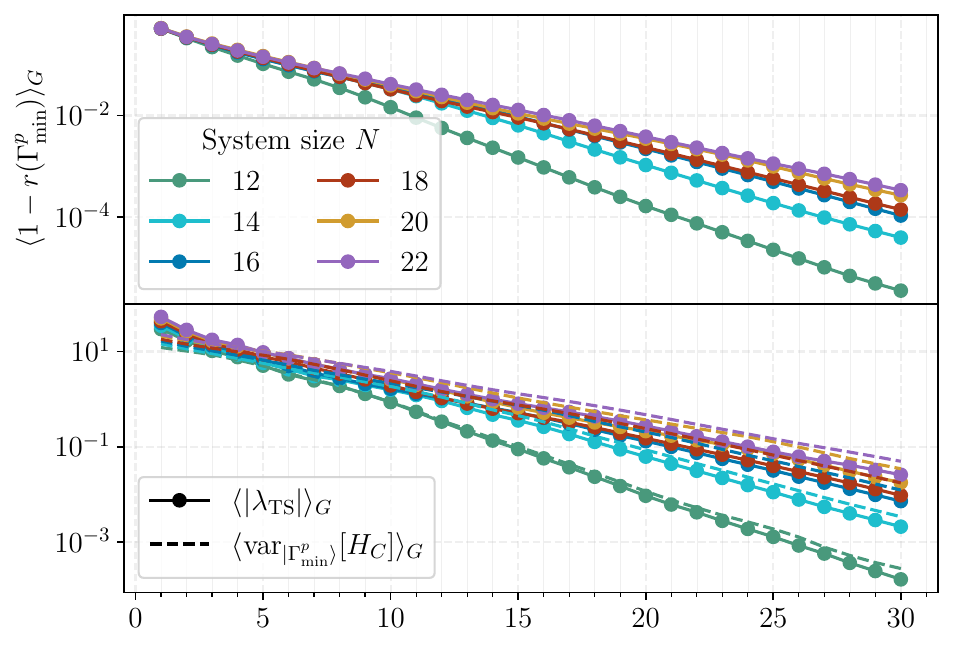}
    \caption{(\textit{Top}) Circuit depth dependence of the approximation ratio $r(\Gamma^p_{\min})$, which approach zero exponentially with $p$. These results were initially observed in~\cite{crooks2018performance, zhou2018quantum}.
    (\textit{Bottom}) Relation between the magnitude of the negative curvature around the transition state $\Gamma^{p+1}_{\mathrm{TS}}$, and the energy variance $\mathrm{var}_{\Gamma^p_{\min}}[H_C]$ as functions of the circuit depth $p$. The numerical data reveals a notable quantitative alignment between the curvature and the energy variance for varying system sizes $N$.}
    \label{fig:curv_and_state_pop}
\end{figure}

Finally, we did not discuss the role of the expectation value, $\langle \phi_0 | H_C - E_0 | \psi_0 \rangle$ in Eq.~(\ref{eq:curv_refined_expr}). On the one hand, this matrix element is expected to be extensive, i.e.~increasing proportionally to the number of degrees of freedom, $N$, as is also confirmed by our numerical simulations, see Figs.~\ref{fig:curv_and_state_pop} and~\ref{fig:curv_and_state_pop_weighted}. On the other hand, estimating the scaling of this matrix element with $p$ remains an open challenge. The contributions to this expectation value primarily arise from eigenstates of $H_C$  where both $|\phi_0\rangle$ and $|\psi_0\rangle$ have significant weight. Developing a framework to accurately assess these contributions and their scaling with $p$ based on physically motivated assumptions remains an intriguing open problem.

Another physical quantity that quantifies the deviation of the state $|\Gamma^p_\text{min}\rangle$ from an eigenstate of the cost Hamiltonian $H_C$ is the energy variance. Using the notation of Eq.~(\ref{eq:expr_for_curv}) we express the energy variance as
\begin{align} \label{Eq:var}
    &\mathrm{var}_{|\Gamma^p_{\min}\rangle}[H_C] = \langle \Gamma^p_{\min}| H^2_C|\Gamma^p_{\min}\rangle - \langle \Gamma^p_{\min}|H_C|\Gamma^p_{\min}\rangle^2 \notag, \\
    &= |\alpha^0_\perp|^2  \mathrm{var}_{|\psi_0\rangle}[H_C] + |\alpha^0_\perp|^2|\alpha^0|^2 (\langle \psi_0| H_C |\psi_0\rangle-E_0 )^2. 
\end{align}
From this expression, it is apparent that the energy variance is proportional to the same infidelity of the state $U(\Gamma^p_{\min})|+\rangle$ to the ground state. 

Comparing expressions~(\ref{eq:curv_refined_expr})-(\ref{Eq:var}), we expect both $|\lambda_{\rm{TS}}|$ and energy variance to display similar qualitative behavior with the circuit depth $p$. Thus, we establish a connection between two seemingly unrelated properties: the energy variance of the quantum state prepared by the QAOA$_{p}$ circuit and the curvature of the cost function of the QAOA$_{p+1}$ at the transition state. Since so far we focused on the ground state $|E_0\rangle$ of $H_C$, and previous literature~\cite{zhou2018quantum, crooks2018performance} heuristically demonstrated that for the \textsc{MaxCut} problem on unweighted 3-regular graphs, the approximation ratio $r(\Gamma)$ tends towards 1 exponentially with increasing circuit depth $p$, we anticipate that infidelity $\alpha^0_\perp$ also tends to zero exponentially, thus leading to an exponential decrease in both curvature and energy variance to zero.
 
To validate our expectations, we first reproduced the numerically observed exponential convergence of QAOA. We then calculated the average absolute value of $\lambda_{\mathrm{TS}}$ across various unweighted \textsc{MaxCut} instances with $N$ ranging from 12 to 22 vertices. The obtained results are displayed in Fig.~\ref{fig:curv_and_state_pop} together with the average of the approximation ratio for circuit depths ranging in the interval $p \in [1, 30]$. The top panel of this figure reproduces the exponential decrease of $1-r$ with circuit depth $p$~\cite{crooks2018performance, zhou2018quantum}. The bottom panel shows the surprisingly close quantitative agreement between the averaged energy variance and the absolute value of the negative curvature $|\lambda_{\mathrm{TS}}|$. This signals that these quantities can be related in a tighter way than what we discussed above, in particular, the constants $\alpha^0_\perp$ and $\kappa^0_\perp$ may be proportional to each other.  

Finally, we want to highlight some possible implications of the above observations for the performance of the QAOA at finite but deep circuit depth. First, our results above imply that if the QAOA prepares an eigenstate $|E_l\rangle$ that is different from the ground state, $l>0$, of the cost Hamiltonian $H_C$ the optimization strategy that uses transition states as initialization will halt since there are no descent directions around any of the transition states. Assuming that the ground state is challenging to prepare for whatever reason, it may be feasible for the QAOA to instead prepare a low-lying eigenstate $|E_l\rangle$ with $l>0$ but not too large, with high fidelity ($|\alpha^l| \sim 1$). As a consequence, the local negative curvature around each TS will become small and as a result, we expect the optimization to slow down. In Appendix~\ref{Sec:App-ts_1}, we illustrate this scenario in an instance of a weighted 3-regular graph that was originally studied in~\cite{zhou2018quantum}. 

Importantly, the exponential decay of the curvature observed in Fig.~\ref{fig:curv_and_state_pop} should not be viewed as a numerical limitation, but rather as a physical signature of the QAOA's convergence to an eigenstate of $H_C$. This is evidenced by the simultaneous exponential decay of both the approximation ratio and the energy variance, with the latter being a direct measure of how close a state is to an eigenstate of $H_C$. As previously shown, when the QAOA prepares an exact eigenstate of $H_C$, both our curvature estimate and the exact Hessian eigenvalue vanish identically. This establishes that the exponential decrease in curvature is a fundamental feature of the algorithm's convergence, rather than a numerical artifact, and is fundamentally different from barren plateau phenomena where small landscape curvature hinders optimization.

Thus, our results that relate the landscape curvature at the TS, the energy variance of the QAOA state, and infidelity, suggest that using the \textsc{Greedy} strategy (or similarly using the TS $\Gamma^{p+1}_{\mathrm{TS}}(1,1)$) the QAOA may effectively converge to a (low) energy manifold of the cost Hamiltonian $H_C$ in the regime of deep circuit. Quantifying how this convergence happens remains an open problem. In the next section, we make the first steps in this direction by expanding the cost function to a higher order around the transition state.

Thus, our results that relate the landscape curvature at the TS, the energy variance of the QAOA state, and infidelity, suggest that using the \textsc{Greedy} strategy (or similarly using the TS $\Gamma^{p+1}_{\mathrm{TS}}(1,1)$) the QAOA may effectively converge to a (low) energy manifold of the cost Hamiltonian $H_C$ in the regime of deep circuit. While a complete characterization of this convergence mechanism remains an open problem, we take initial steps toward understanding it in the next section through a higher-order expansion of the cost function around the transition state.

\section{Higher order expansion of energy along index-1 direction \label{Sec:IV}}
In this section we use the approximate index-1 direction given by Eq.~\eqref{Eq:negative} to expand the QAOA$_{p+1}$ cost function up to the fourth order along the descent direction. This results in lower bound on the improvement of the QAOA cost function resulting from increasing the number of parameters from $2p$ to $2p+2$.

\subsection{Taylor expansion}
Using the explicit knowledge of the index-1 descent direction, we estimate how much the energy can be improved using the Taylor series expansion around the point $\Gamma_\mathrm{TS}^{p+1}$. Specifically, Appendix~\ref{Sec:App-expand} details our computation of the cost function's expansion, $E(\Gamma_{\mathrm{TS}}^{p+1} +\varepsilon {\bm \delta}_{\mathrm{TS}})$, to the \emph{fourth order} in $\varepsilon$. In what follows, we neglect the cubic term in the energy expansion, delegating the details of such step to the Appendix~\ref{Sec:App-expand}, and obtain a simple expression:
\begin{equation}\label{Eq:E-expand}
E(\Gamma_\text{TS}^{p+1} +\varepsilon  {\bm \delta}_\text{TS} )
\approx 
 E(\Gamma^p_\text{min}) + \frac{\lambda_{\rm{TS}}}{2} \varepsilon^2 + \partial_{\gamma_1}^2 E(\Gamma^{p+1}_{\text{TS}}) \varepsilon^4,
\end{equation} 
where $\partial_{\gamma_1}^2 E(\Gamma^{p+1}_{\text{TS}})$ is expressed as a combination of two expectation values:
\begin{multline} 
 \label{Eq:c}
\partial^2_{\gamma_1} E(\Gamma^{p+1}_{\rm{TS}})= 
2\langle +|H_C U(\Gamma^p_\text{min})^\dagger H_C U(\Gamma^p_\text{min}) H_C|+\rangle 
\\
- 2\Re{\langle +|U(\Gamma^p_\text{min})^\dagger H_C U(\Gamma^p_\text{min}) H_C^2|+\rangle}.
\end{multline}
The fact that the fourth order expansion term of energy is proportional to the second derivative, $ \partial^2_{\gamma_1} E(\Gamma^{p+1}_{\rm{TS}})$ can be understood from the specific form of the descent direction vector, Eq.~\eqref{Eq:descent}. Using the explicit form of the descent vector, in Appendix~\ref{Sec:App-expand} we show that the first non-trivial expansion term in $\varepsilon$ of the state $U(\Gamma^{p+1}_\text{TS}+\varepsilon {\bm \delta}_\text{TS})\ket{+}$ is proportional to $\varepsilon^2 H_C\ket{+}$. Combining two of such terms, we precisely get the contribution in the first line of Eq.~(\ref{Eq:c}) above, which is thus coming with an order of $\varepsilon^4$. 

\begin{figure}
    \centering
    \includegraphics[width=0.99\linewidth]{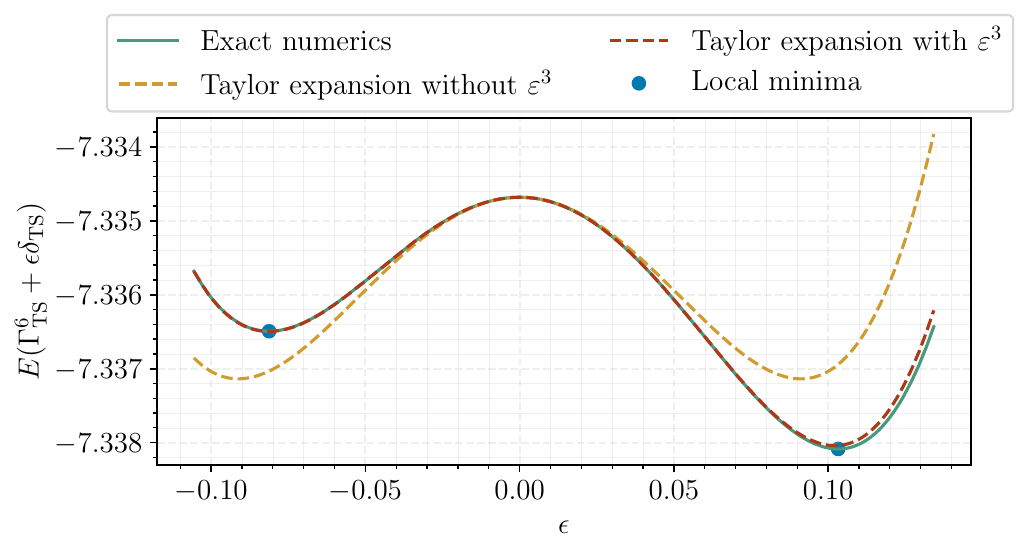}
    \caption{Taylor approximation of the energy at a transition state obtained from a local minima of QAOA$_5$ when perturbed in the index-1 direction. We inspect the impact of the cubic term in the perturbation parameter $\varepsilon$ in the energy expansion around the index-1 direction. The instance studied corresponds to that of Appendix~\ref{Sec:App-ts_1}.} 
    \label{fig:taylor_energy}
\end{figure}

In Fig.~\ref{fig:taylor_energy}, we assess the Taylor approximation's accuracy (which incorporates a cubic term in $\varepsilon$) against exact numerical data obtained by computing the Hessian at $\Gamma^{p+1}_{\mathrm{TS}}$ and determining the energy along the exact index-1 direction. Additionally, we examine the impact of omitting the cubic term from the expansion. While the omission of the cubic term leads to underestimation of the energy improvement, it significantly streamlines the energy expansion analysis and still faithfully replicates the qualitative behavior of exact energy depenence on the slice.

\subsection{Comparing estimated and true energy gains\label{sec:iv_B}}
Using the expression for the energy Eq.~\eqref{Eq:E-expand} along the index-1 direction we compute the value of the perturbation parameter $\varepsilon$ that minimizes the energy in this univariate optimization problem. The solution then reads
\begin{multline}
\label{eq:energy_delta_final}
\Delta E(\varepsilon_*)=E(\Gamma^{p+1}_{\text{TS}} + \varepsilon_* \delta_{\text{TS}})- E(\Gamma^p_{\min})  \\
=-\frac{\lambda_{\mathrm{TS}}^2}{16 \partial^2_{\gamma_1}E(\Gamma^{p+1}_{\mathrm{TS}})},
\end{multline}
with the distance from the transition state to the local minimum on the slice corresponding to $\varepsilon^2_* = -\lambda_{\mathrm{TS}}/4\partial^2_{\gamma_1}E(\Gamma^{p+1}_{\mathrm{TS}})$. We note that although from the expansion of the cost function we get the local energy minimum, this is the artifact of the considering only one out of $2p+2$ directions in the energy landscape. When viewed without projection, we do not expect the point $\Gamma^{p+1}_{\text{TS}} + \varepsilon_* \delta_{\text{TS}}$ to be a local minimum or even a saddle point of the cost function, see Fig.~\ref{Fig:cartoon}(b). 

Using the expression for $\lambda_\text{TS}$ obtained in the previous section in Eq.~\eqref{eq:curv_refined_expr} we see that the numerator in Eq.~\eqref{eq:energy_delta_final} is proportional to the square root of the infidelity $\alpha^0_\perp$ to the ground state $|E_0\rangle$. 
Furthermore, we expect that $\partial^2_{\gamma_1}E(\Gamma^{p+1}_{\mathrm{TS}})$ in the denominator remains finite and extensive in the deep QAOA limit. In particular, in Appendix~\ref{Sec:App-expand} we discuss that $\partial^2_{\gamma_1}E(\Gamma^{p+1}_{\mathrm{TS}})$ can be approximated as follows:
\begin{align}
    \label{eq:c_approx}
    \partial^2_{\gamma_1}E(\Gamma^{p+1}_{\mathrm{TS}}) &\approx 2 n_C^2 \langle +|  \frac{H_C}{n_C} U^\dagger(\Gamma^p_{\min}) H_C U(\Gamma^p_{\min}) \frac{H_C}{n_C}  |+\rangle \notag \\ &- 2n_C^2 E(\Gamma^p_{\min}).
\end{align}
where for clarity we explicitly singled out the common factor of $n_C$ that highlights the scaling of quartic expansion coefficient. From Eq.~(\ref{eq:c_approx}) the quartic coefficient in the expansion of energy, $\partial^2_{\gamma_1}E(\Gamma^{p+1}_{\mathrm{TS}})$, can be understood as the energy difference between states $U(\Gamma^p_{\min})H_C|+\rangle$ and $U(\Gamma^p_{\min})|+\rangle$, multiplied by the extensive constant $n_C^2 \sim N$. It is natural to expect that QAOA circuit, when applied to the excited eigenstate of the mixing Hamiltonian, $H_C|+\rangle$, yields the final state that has higher energy compared to the state $U(\Gamma^p_{\min})|+\rangle$, that QAOA circuit by design aims to rotate into the ground state of classical Hamiltonian. This physical reasoning implies that the quartic expansion coefficient, $\partial^2_{\gamma_1}E(\Gamma^{p+1}_{\mathrm{TS}})$, is positive, as is also confirmed in numerical simulations. As the algorithm converges at large enough circuit depths $p$, we expect $\partial^2_{\gamma_1}E(\Gamma^{p+1}_{\mathrm{TS}})$ to plateau at an extensive value. Finally, it is important to note that the $n_C^2$ factor in the denominator of Eq.~\eqref{eq:energy_delta_final} cancels out with the same factor coming from $|\lambda_{\mathrm{TS}}| \propto n_C$, see Eq.~(\ref{eq:curv_refined_expr}).

\begin{figure}[t]
    \centering
    \includegraphics[width=0.99\linewidth]{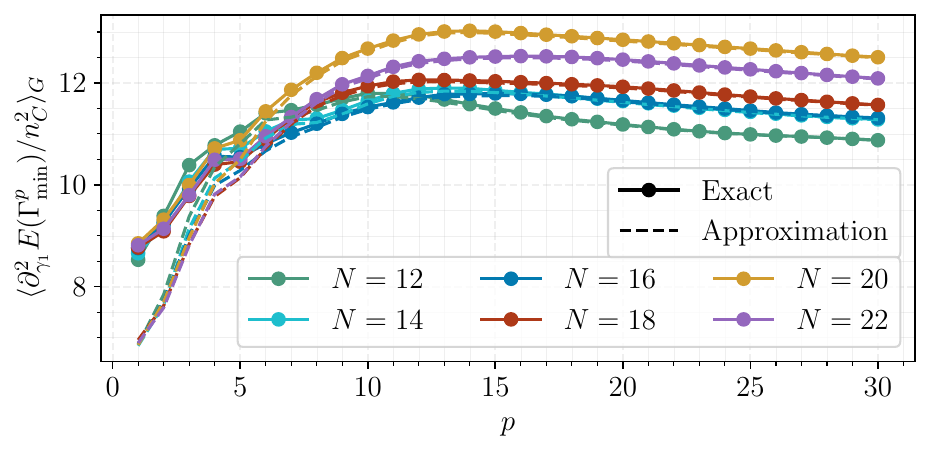}
    \caption{Averaged circuit depth behavior of $\partial^2_{\gamma_1}E(\Gamma^{p+1}_{\mathrm{TS}})$ and its approximation Eq.~\eqref{eq:c_approx} for different system sizes agree for $p\geq 5$.}
    \label{fig:c_averaged}
\end{figure}

We use numerical simulations to verify the validity of the approximation given by Eq.~\eqref{eq:c_approx}, on random instances of unweighted 3-regular graphs with $N=[12, 22]$ vertices. From Fig.~\ref{fig:c_averaged}, we observe a clear extensive behavior of $\partial^2_{\gamma_1}E(\Gamma^{p+1}_{\mathrm{TS}})$. Interestingly, we see that even though the data for different system sizes does not perfectly collapse onto a single curve, its dependence on the system size at all circuit depths is relatively weak. For example, at $p=30$ the values for different systems sizes lie in the interval $[11, 13]$.

Using the intuition that the quartic term that represents the denominator in the expression for energy gain, Eq.~\eqref{eq:energy_delta_final} is saturating to the finite value for large $p$, we conclude that the lower bound on the energy gain is proportional to the curvature around the transition state, and thus is expected to decrease exponentially with the circuit depth $p$, as supported by the analysis in the previous section.  We numerically check the tightness of the lower bound provided by Eq.~\eqref{eq:energy_delta_final}.

\begin{figure}
    \centering
    \includegraphics[width=0.99\linewidth]{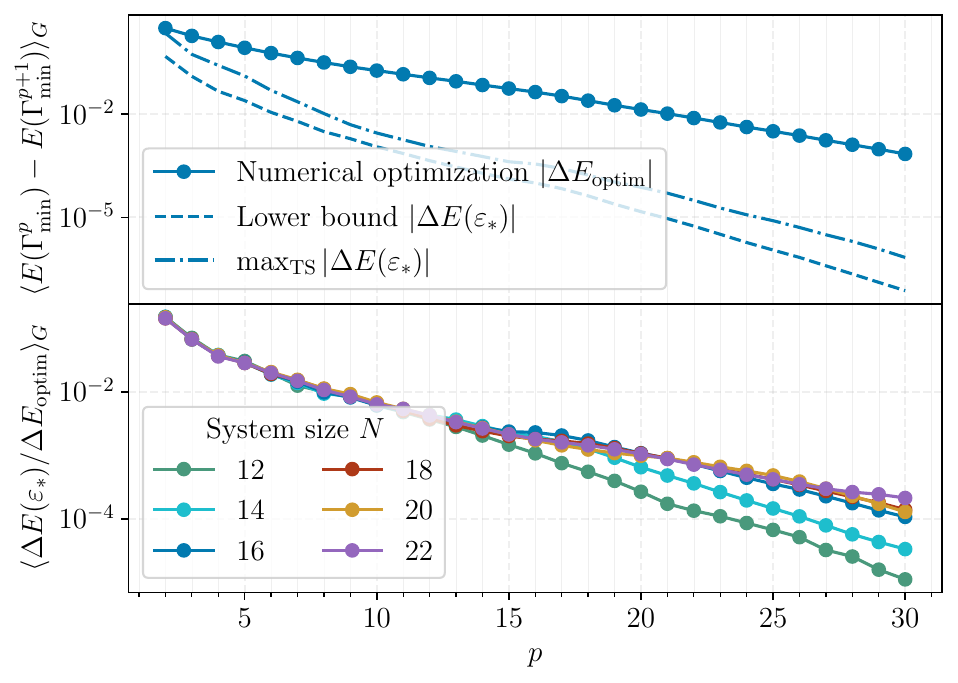}
    \caption{(\textit{Top}) Average energy improvement between local minima of QAOA$_{p}$ and QAOA$_{p+1}$ as a function of the circuit depth $p$ for an unweighted 3-regular graph with $N=16$ vertices. The lower bound Eq.~\eqref{eq:energy_delta_final}, which relies on local information about the cost function landscape around index-1 saddle points overestimates the results obtained by numerically optimizing using the \textsc{Greedy} strategy of~\cite{Sack2022}. (\textit{Bottom}) Averaged quality of the lower bound on the energy improvement, as given by $\Delta E(\varepsilon_*)/\Delta E_{\mathrm{optim}}$, for systems sizes ranging from 12 to 22 vertices.}
    \label{fig:eslice_vs_true}
\end{figure}

To this end, in Fig.~\ref{fig:eslice_vs_true} we first take random instances of unweighted 3-regular graphs with $N=16$ vertices and compare Eq.~\eqref{eq:energy_delta_final} to the energy improvement coming from performing numerical optimization, using the Broyden–Fletcher–Goldfarb–Shanno (BFGS) algorithm~\cite{bfgs1, bfgs2, bfgs3, bfgs4}, following the \textsc{Greedy} strategy introduced in~\cite{Sack2022}. We also show the best improvement selected from improvements obtained from moving along the index-1 direction of $2p+1$ distinct TS obtained from the initial local minima $\Gamma^p_{\min}$, labeled as $\max_\text{TS}[\Delta E(\varepsilon_*)] = \max_{(i,j) }\big[E(\Gamma^p_{\min})-E(\Gamma^{p+1}_{\mathrm{TS}}(i,j) + \varepsilon \delta_{\mathrm{TS}})\big]$. 

Figure~\ref{fig:eslice_vs_true} reveals that the true energy improvement, and our lower bound Eq.~\eqref{eq:energy_delta_final} both decrease exponentially with $p$, although with different slopes. In particular, the energy improvement from moving alongside the index-1 direction underestimates the improvement obtained from numerical optimization. This allows us to conclude that although the BFGS optimization algorithm using a large number of iterations can find better local minima by moving far away from the transition state, the entire cost function landscape is getting more ``flat'' with $p$. Thus, while our lower bound, which is conceptually similar to one step of local optimization, is underestimating the magnitude of the energy improvement, it has the same functional dependence on $p$. It remains to be understood if one can establish a (heuristic) relation between our bound and the true energy decrease, thus allowing us to predict the QAOA performance quantitatively from Eq.~\eqref{eq:energy_delta_final}.

Finally, we discuss the scaling of energy improvement with system size, as shown in the bottom panel of Fig.~\ref{fig:eslice_vs_true}. Similar scaling is also evident in instances of \textsc{MaxCut} on 3-regular weighted graphs, as illustrated in Fig.~\ref{fig:eslice_vs_true_weighted} in Appendix~\ref{App:numerics}. The numerical results show that at a fixed circuit depth our bound on energy improvement is proportional to the system size $N$. Indeed, we show that the ratio between the numerical energy improvement and our estimate tends to a constant value with increasing system size, and numerical energy improvement is known to be proportional to $N$. Analytically, this behavior arises from the expectation value $|\langle \phi_0| H_C - E_0 |\psi_0\rangle|$ in the expression for the approximate negative Hessian eigenvalue in Eq.~\eqref{eq:curv_refined_expr} which is extensive in $N$. In summary, the scaling of our bound on energy improvement with $N$ implies that the improvement in approximation ratio does not scale with 
$N$, which is consistent with the gains from numerical optimization.

\section{Discussion\label{Sec:V}}
In this work, we perform an analytic study of transition states of the QAOA cost function, that were constructed in Ref.~\cite{Sack2022}.  
These transition states are characterized by the vanishing gradient of the cost function and a unique negative eigenvalue of Hessian. In the present work, we provide an accurate \emph{analytic} estimate of the minimum eigenvalue of the Hessian and its corresponding eigenvector for each of the $2p+1$ TS. Moreover, we relate the curvature in the vicinity of transition states to physical observables such as the infidelity of the ground state preparation of the QAOA circuit, and construct the higher-order expansion of the QAOA cost function along the negative curvature direction, which allows us to put a lower bound on the QAOA cost function improvement. 

Crucially, the results obtained in this paper are recursive. Assuming that QAOA at depth $p$ found a local minimum, we provide a lower bound on the cost function improvement of the QAOA at depth $p+1$. Thus, our approach is applicable to QAOA in the regime of large $p$ and we envision that it may be potentially used to obtain a QAOA performance guarantee~\cite{farhi2014quantum,wurtz2021maxcut, farhi2020wholegraph}. Indeed, the only missing link in such performance guarantee remains to be the bound on the improvement in infidelity, which determines the landscape curvature and lower-bounds energy improvement as shown by our work. 

Beyond being a potential step towards performance guarantee, the significance of these estimates is twofold: first, they substantially reduce the computational effort required to implement the \textsc{Greedy} optimization strategy outlined by~\cite{Sack2022}, as they circumvent the need to construct and diagonalize the Hessian of the cost function at each TS. Second, our results establish the analytical framework that reveals properties of the QAOA cost function at arbitrarily large $p$. 

In particular, we show that for unweighted 3-regular graphs, the negative curvature of the landscape at the TS is intimately connected to the energy variance of the QAOA state. This not only offers a physical interpretation of the negative curvature at the TS but also raises questions about the QAOA performance in the limit of large circuit depth. While it is anticipated that the QAOA will prepare the ground state as $p \to \infty$~\cite{farhi2014quantum}, our findings suggest that the QAOA may effectively converge to a (low) energy manifold of the cost Hamiltonian $H_C$ in the deep circuit regime.

Furthermore, our numerical analyses indicate that the lower bound on the energy improvement has the same qualitative dependence on the QAOA depth as the true energy improvement. At the same time, our lower bound parametrically underestimates the actual improvements achieved through numerical optimization. This observation suggests that the local vicinity of the transition state that we can explore analytically may be non-trivially related to the global properties of the QAOA cost function. Establishing such a relation even heuristically may be useful for accurately forecasting the performance of the QAOA, and may provide a useful step towards a more complete understanding of the QAOA.

Finally, given the broad applicability of the transition states-based approach, it becomes intriguing to consider its extension to problems beyond the \textsc{MaxCut}. Exploring the impact of different cost and mixer Hamiltonians on the QAOA cost function landscape presents a promising avenue for future research. Additionally, applying the transition state (TS) strategy to other variational algorithms offers an exciting opportunity. By leveraging the unique characteristics of both the circuit and the problem structure, it may be possible to provide similar initialization strategies and devise analytic estimates for the improvement of the cost function from the optimization process.

\begin{acknowledgements}
We acknowledge S.~Sack and R.~Kueng for useful discussions and previous collaborations on Ref.~\cite{Sack2022}. We also acknowledge useful discussion with P.~Brighi, M.~Ljubotina, A.~Michailidis, G.~Matos, V.~Karle, A.~Kerschbaumer, D.~Kolisnyk, and M.~S.~Rudolph and thank P.~Brighi, S.~Sack, V.~Karle, A.~Kerschbaumer, A. Montanaro, and L. Zhou for their valuable feedback on the manuscript.
We acknowledge support by the European Research Council (ERC) under the European Union's Horizon 2020 research and innovation program (Grant Agreement No.~850899).
\end{acknowledgements}

\appendix
\section{Numerical simulations\label{Sec:App-ts_1}}\label{App:numerics}
In this appendix, we first introduce the package used for numerical simulations throughout this work. Next, we justify the choice of the first transition state as a focus of the main paper. For this, we demonstrate that considering only the first transition state, defined in Eq.~\eqref{Eq:descent}, does not degrade the initialization performance compared to the case when one uses $2p+1$ TS. Next, we discuss and show how the QAOA  can converge to an excited state of the cost Hamiltonian $H_C$ for a wide range of circuit depths $p$, and how this is reflected in the non-monotonic behavior of the landscape curvature at the transition state. Finally, we conclude this appendix by discussing numerical results for weighted instances of \textsc{MaxCut} on 3-regular graphs. 

\subsection{Package \texttt{QAOALandscapes.jl} for numerical simulations of QAOA in Julia}
All the simulations performed in this work were conducted using the Julia programming language~\cite{julia} and the package \texttt{QAOALandscapes.jl}\cite{qaoa_landscapes}, which was developed by one of the authors. This package is designed to apply the QAOA to solve general combinatorial optimization problems by encoding the classical problem into a $k$-spin classical Hamiltonian. It relies on matrix-free operations to enhance speed and reduce memory usage. Currently, it supports only the Pauli-$X$ mixer operator (see Eq.\eqref{Eq:mixing}); however, additional mixers can be incorporated as is described in the documentation. The package includes support for both CPU and GPU backends, with GPU capabilities for CUDA and Metal-based devices through \texttt{CUDA.jl}~\cite{besard2018juliagpu, besard2019prototyping} and \texttt{Metal.jl}~\cite{metal_jl} packages respectively.

Numerical optimization was performed using the \texttt{Optim.jl}~\cite{optim, optim_2} package, and in particular using the BFGS~\cite{bfgs1, bfgs2, bfgs3, bfgs4} algorithm. For this, the package supports fast and exact gradient calculations through the use of automatic differentiation using the method introduced in~\cite{yao, jones_gradient}. Further details on how to use \texttt{QAOALandscapes.jl} can be found on the Readme and documentation in~\cite{qaoa_landscapes}.

\subsection{Quality of optimization using only $\Gamma^{p+1}_{\mathrm{TS}}(1,1)$}
The \textsc{Greedy} approach introduced in~\cite{Sack2022}, requires one to launch optimization twice from each of the $2p+1$ TS constructed from a local minimum $\Gamma^p_{\min}$ of QAOA$_p$. The need to launch optimization from $2p+1$ distinct transition states and moving in two potential directions away from the saddle accounts for a $2(2p+1)$  overhead on top of heuristic initializations like \textsc{Interp} and \textsc{Fourier}~\cite{zhou2018quantum} with similar performance. Even though \textsc{Greedy} comes with a guarantee of improvement at each circuit depth $p$, it is desirable to further reduce the optimization cost that it incurs. 

\begin{figure}
    \centering
    \includegraphics[width=0.99\linewidth]{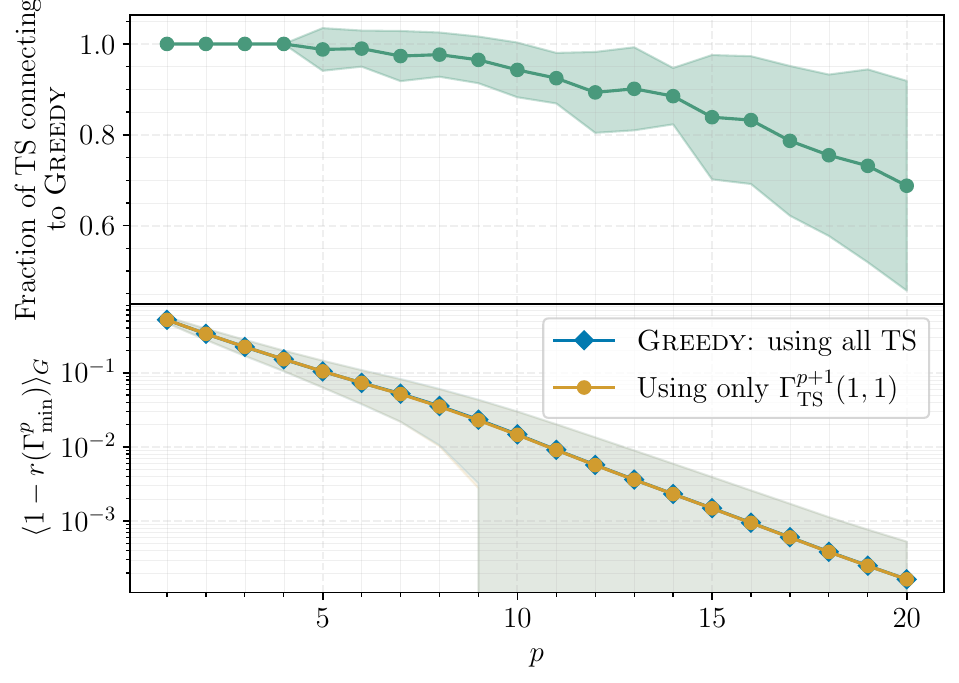}
    \caption{(\textit{Top}) Fraction of the $2p+1$ TS constructed from a local minima $\Gamma^p_{\min}$ that connect to the \textsc{Greedy} solution. The data corresponds to instances of random 3-regular unweighted graphs with $N=12$ vertices. (\textit{Bottom}) Performance of numerical optimization using only the transition state with zeros padded at indices $(\beta, \gamma)=(1,1)$. The average performance, over instances of 3-regular unweighted graphs with $N=12$ vertices seems effectively identical to that of the \textsc{Greedy} strategy~\cite{Sack2022} that uses the set of all $2p+1$ TS constructed from a local minima of QAOA$_p$.}
    \label{fig:ts_to_greedy}
\end{figure}

We thus inspect given a local minimum $\Gamma^p_{\min}$ what fraction of the $2p+1$ TS constructed from it leads to the \textsc{Greedy} solution after optimization. We numerically observe in Fig.~\ref{fig:ts_to_greedy} that the number of TS that connect through optimization to the best local solution decreases with $p$ but remains finite at approximately 0.7 at circuit depth $p=20$. In all cases, we observed that the transition state constructed by padding with zeros the first layers, i.e. $\Gamma^p_{\mathrm{TS}}(1,1)$ in the notation of~\cite{Sack2022}, leads to the \textsc{Greedy} solution as also shown in the figure, where we plot the average approximation ratio obtained from using the \textsc{Greedy} strategy, and the result coming from only using $\Gamma^p_{\mathrm{TS}}(1,1)$. 

This observation motivates us to focus on the transition state $\Gamma^p_{\mathrm{TS}}(1,1)$ as it provides a reliable choice of an initial transition state. 
Fixing the transition state in such a way reduces the cost of optimization to a simple factor of two, while still keeping the guarantee of improvement. Furthermore, as we will show below, using $\Gamma^p_{\mathrm{TS}}(1,1)$ enables us to obtain a lower bound on the energy improvement of the QAOA between circuit consecutive circuit depths. 

\subsection{Converging to an excited state}
\begin{figure}[t]
    \centering
    \includegraphics[width=0.99\linewidth]{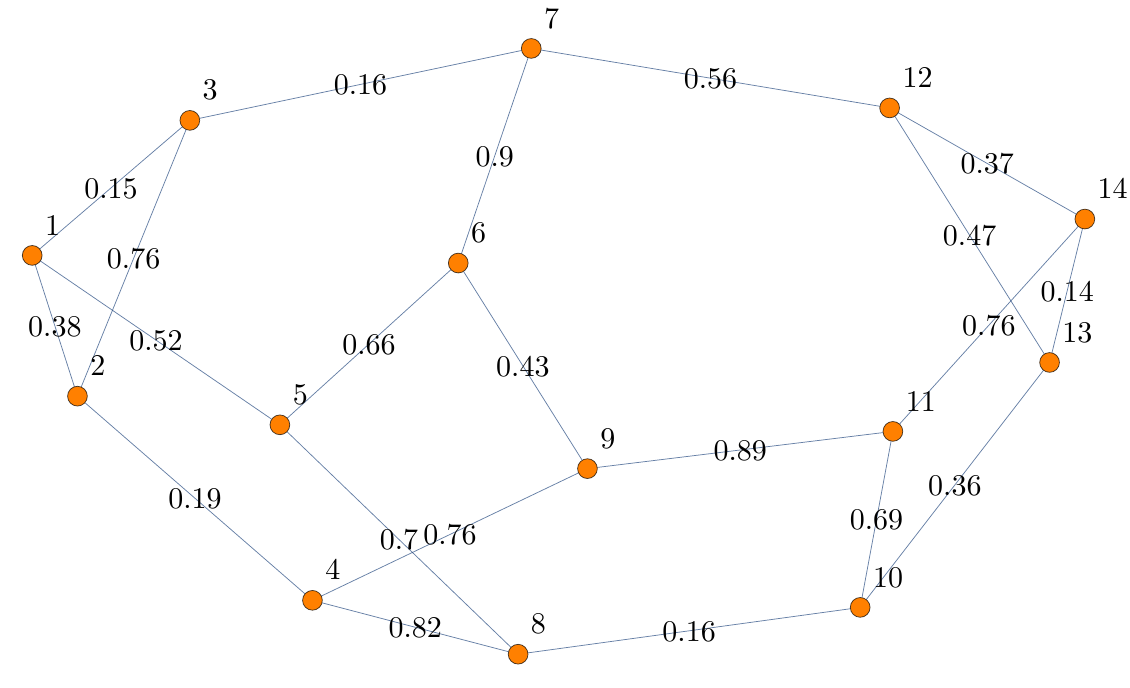}
    \caption{Instance of \textsc{MaxCut} with $N=14$ vertices where the QAOA algorithm gets trapped in local optima, and mostly converges to the first excited state of the cost Hamiltonian $H_C$.}
    \label{fig:harvard_instance}
\end{figure}

Here we provide a specific example of QAOA converging to an excited state. 
To this end, we use a \textsc{MaxCut} instance studied in Ref.~\cite{zhou2018quantum}. 
This instance was used to highlight the performance of the \textsc{Fourier} and \textsc{Interp} strategies~\cite{zhou2018quantum} on ``hard" instances of \textsc{MaxCut}. The authors used the minimum spectral gap $\Delta_{\min}$ of the annealing Hamiltonian
$$
H_{\mathrm{QA}}(s) = s H_C + (1-s) H_B; \; s \in [0, 1]
$$
to distinguish between ``hard" and ``easy" instances for optimization. In particular, for the instance shown in Fig.~\ref{fig:harvard_instance} one can show that $\Delta_{\min} < 10^{-3}$, which translates into prohibitively long annealing time~\cite{zhou2018quantum}. 

Both \textsc{Fourier} and \textsc{Interp} initialization strategies reuse a local minimum of the QAOA at circuit depth $p$ to construct an initialization at circuit depth $p+1$. In the case of the \textsc{Fourier} initialization --- which will also use here --- the idea is to use a different parametrization of QAOA. Instead of using the $2p$-parameters $(\bm{\beta}, \bm{\gamma})$, Ref.~\cite{zhou2018quantum} considers the discrete cosine and sine transform of $\bm{\beta}$ and $\bm{\gamma}$ respectively
\begin{align}
\beta_i &= \sum_{k=1}^q v_k \cos \big[ (k-\frac{1}{2}) (i-\frac{1}{2})\frac{\pi}{p} \big], \\
\gamma_i &= \sum_{k=1}^q v_k \sin \big[ (k-\frac{1}{2}) (i-\frac{1}{2})\frac{\pi}{p} \big].
\end{align}
Through such coordinate transformation, the new parameters become the amplitudes $(\bm{v}, \bm{u})$ of the frequency components for $\bm{\beta}$ and $\bm{\gamma}$, respectively. The basic \textsc{Fourier}$[\infty, 0]$ variant of the strategy, generates a good initial point for  QAOA$_{p+1}$ by adding a higher frequency component, initialized at zero amplitude, to the optimum at level $p$. Last, in the improved variant \textsc{Fourier}$[\infty, R]$ in addition to optimizing according to the basic strategy, we optimize QAOA$_{p+1}$ from
$R+1$ extra initial points, $R$ of which are generated by
adding random perturbations to the best of all local optima $(\bm{v} , \bm{u})$ found at level $p$ (see the Appendix B.2 in Ref.~\cite{zhou2018quantum} for more details). It is crucial to note that the number of random perturbations in Fourier space, denoted by $R$, serves as a hyperparameter; optimizing its value is essential for enhanced convergence, requiring multiple runs of the optimization process with varied $R$ settings to determine the most effective configuration.  
\begin{figure}
    \centering
    \includegraphics[width=0.99\linewidth]{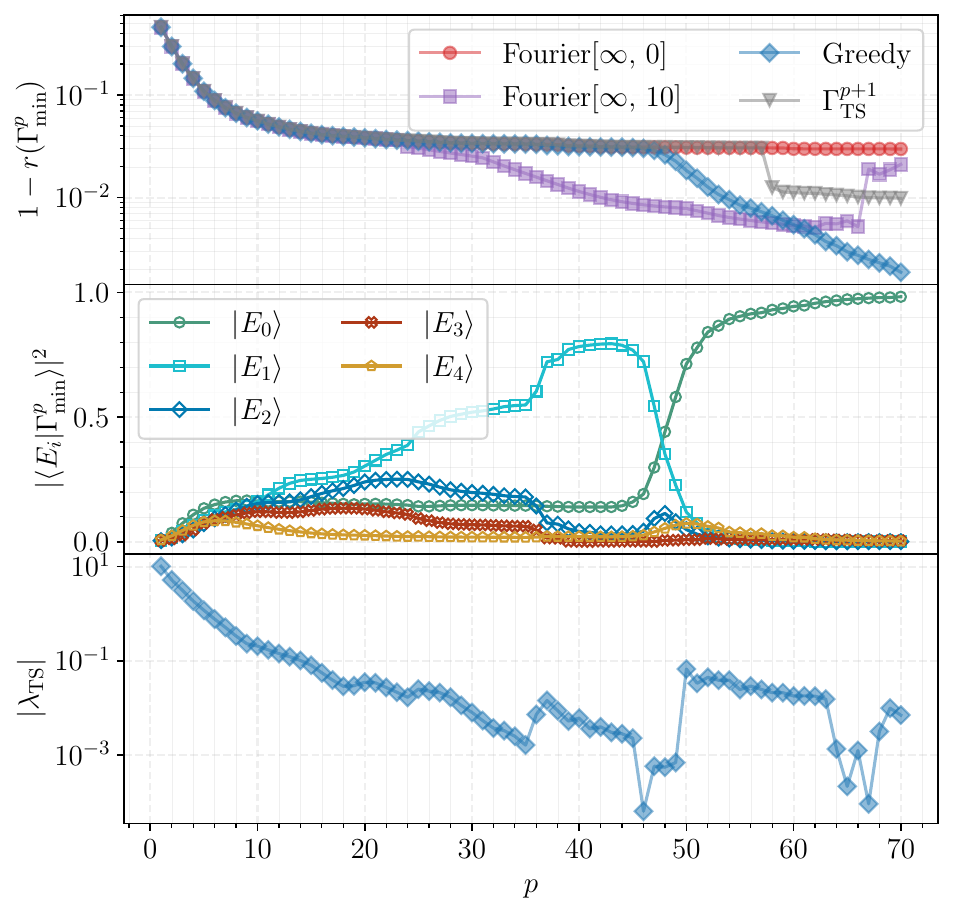}
    \caption{(\textit{Top}) Behavior of the approximation ratio as a function of the circuit depth, for different optimization strategies. (\textit{Middle)} Probability of measuring the fifth lowest energy eigenstates as a function of the circuit depth for the \textsc{Greedy} strategy. The ground state population remains unchanged for a wide range of circuit depths, followed by a sudden increase which correlates with the QAOA overcoming local minima. (\textit{Bottom}) Circuit depth dependence of the landscape curvature at the transition state defined in Eq.~\eqref{Eq:descent} following the \textsc{Greedy} strategy. The curvature displays a gradual decrease, followed by a significant increase when the QAOA overcomes local minima. }
    \label{fig:harvard_perf}
\end{figure}

In Fig.~\ref{fig:harvard_perf} we compare the performance of the QAOA under the \textsc{Fourier}$[\infty, 0]$, \textsc{Fourier}$[\infty, 10]$, \textsc{Greedy}, and  $\Gamma^{p+1}_{\mathrm{TS}}$ strategies. From the approximation ratio, we note that all strategies yield the same performance for circuit depths $p\in [1, 24]$, yet the improvement of the approximation ratio is stalled for $p\geq 10$. We attribute this behavior to the fact that QAOA prepares an excited state of the system with progressively increased fidelity, see the middle panel of Fig.~\ref{fig:harvard_perf}. Eventually, however, the QAOA is able to escape the local minimum that prepares an excited state and starts converging to the ground state.

When the QAOA escapes the local minimum that prepares an excited state of the classical Hamiltonian depends on the initialization scheme. The initialization \textsc{Fourier}$[\infty, 10]$ is the first to escape local optima at $p\sim 24$. \textsc{Greedy} follows next and only manages to escape around $p\sim 46$. Despite this, we observe that the final approximation is consistently better in \textsc{Greedy} than in all other strategies. Interestingly, we see that using only the first transition state as an initialization yields worse performance than \textsc{Greedy} and does escape the local minima that prepares the excited state but at circuit depths $p\sim 58$. Moreover, the initial variant of the \textsc{Fourier} strategy fails to escape from local optima and gets stalled for all circuit depths explored.  

Following our comparative analysis, we articulate two critical observations. First, we conclude that QAOA is capable of preparing low-lying excited states of the classical Hamiltonian. How soon QAOA escapes from such a trap depends on the initialization scheme used, but this phenomenon is present for all initialization routines considered here. Second, we conclude that the landscape curvature at the initial transition state \(\Gamma^{p+1}_{\mathrm{TS}}\), quantified in Eq.~\eqref{eq:curv_first_ts}, is a good indicator of such QAOA behavior. Indeed, the bottom panel in Fig.~\ref{fig:harvard_perf} demonstrates that although the approximation ratio is stalling, the curvature keeps decreasing when QAOA prepares the excited state with progressively higher fidelity. As soon as QAOA starts converging to the true ground state, the curvature shows an increase and then continues to reduce. 

All in all, our results suggest that there may be scenarios, particularly at large system sizes $N$, where the QAOA experiences stagnation, with negligible performance gains across a wide range of circuit depths $p$. This stagnation primarily arises because the QAOA effectively converges to a low-energy manifold of $H_C$, characterized by a small landscape curvature. In the context evaluated in this study Fig.~\ref{fig:harvard_instance}, this manifold principally consists of the ground state and the first excited state Fig.~\ref{fig:harvard_perf}. Our results suggest that the curvature provides useful insights into the QAOA behavior, complementary to the behavior of the approximation ratio. 

\begin{figure}[t]
    \centering
    \includegraphics[width=0.99\columnwidth]{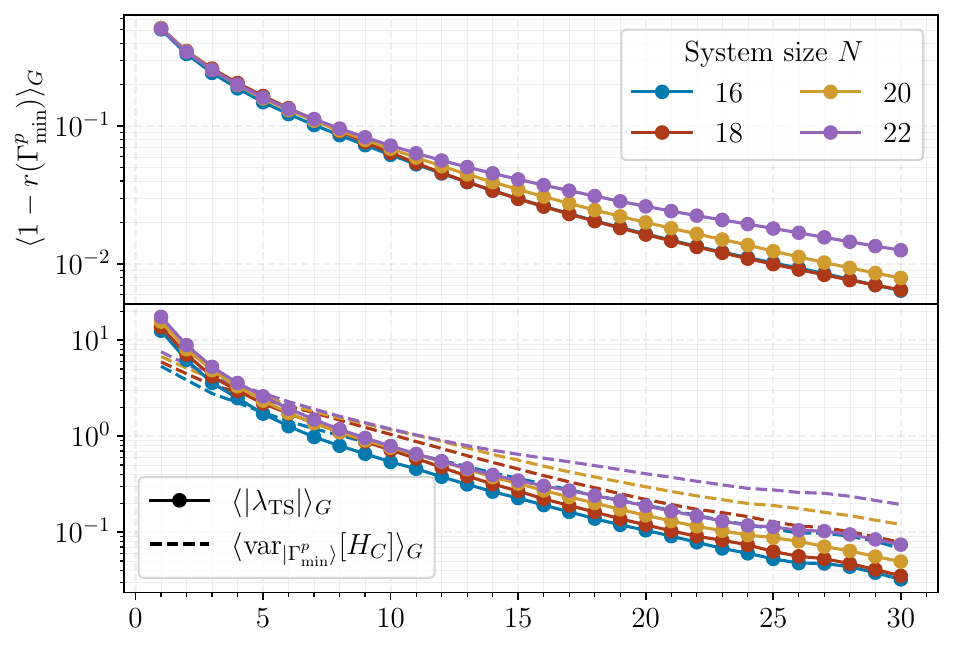}
    \caption{(\textit{Top}) Circuit depth dependence of the approximation ratio $r(\Gamma^p_{\min})$. The scaling of the approximation ratio with the circuit depth $p$ matches the numerical results from~\cite{zhou2018quantum}.
    (\textit{Bottom}) Relationship between the magnitude of the negative curvature around the transition state $\Gamma^{p+1}{\mathrm{TS}}$ and the energy variance $\mathrm{var}_{\Gamma^p_{\min}}[H_C]$ as functions of circuit depth $p$. Although there appears to be qualitative agreement between the curvature and the energy variance across varying system sizes $N$, it is not as close as for the unweighted instances.}
    \label{fig:curv_and_state_pop_weighted}
\end{figure}

\subsection{Numerical results for weighted 3-regular graphs \label{app:extra_numerical_res}}
In this section we present numerical results analogous to those shown in the main text, but for weighted instances of \textsc{MaxCut} on 3-regular graphs.

We start by examining the curvature of the QAOA energy landscape at the transition state from Eq.\eqref{Eq:descent}, alongside the variance of the cost Hamiltonian in the QAOA state. In Fig.~\ref{fig:curv_and_state_pop_weighted} we notice behavior similar to that for unweighted instances described in the main text, albeit with notable differences. First,  the approximation ratio decays slower with the system size, aligning with findings from previous studies on similar \textsc{MaxCut} instances~\cite{zhou2018quantum}. Second, while there is a close qualitative relationship between the energy variance of the QAOA state and the landscape curvature at the transition state in Eq.~\eqref{Eq:descent}, this relationship is not as accurate as for unweighted instances, and energy variance decays slower compared to the curvature with QAOA depth $p$.

Finally, we check the accuracy of the lower bound on energy improvement from Eq.~\eqref{eq:energy_delta_final}. Similar to observations with unweighted \textsc{MaxCut} instances, the lower bound significantly underestimates the energy improvement achieved by QAOA between consecutive circuit depths. At the same time, the ratio between true energy improvement and our lower bound seems to be a universal function of $p$ across a broad range of system sizes considered here.  

\begin{figure}[t]
    \centering
    \includegraphics[width=0.99\columnwidth]{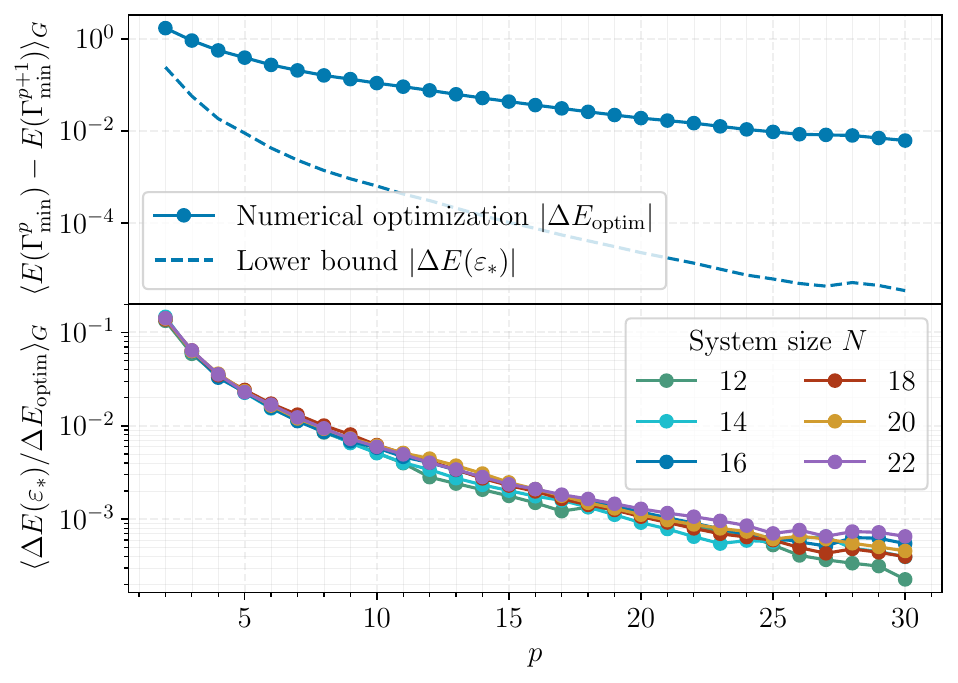}
    \caption{(\textit{Top}) Average energy improvement between local minima of QAOA$_{p}$ and QAOA$_{p+1}$ as a function of the circuit depth $p$. The lower bound Eq.~\eqref{eq:energy_delta_final}, which relies on local information about the cost function landscape around index-1 saddle points overestimates the results obtained by numerically optimizing using the \textsc{Greedy} strategy of~\cite{Sack2022}. (\textit{Bottom}) Averaged quality of the lower bound on the energy improvement, as given by $\Delta E(\varepsilon_*)/\Delta E_{\mathrm{optim}}$, for systems sizes ranging from 12 to 22 vertices.}
    \label{fig:eslice_vs_true_weighted}
\end{figure}

\section{Bounds on the Hessian eigenvalue and eigenvector. \label{Sec:App-bound}}

In this Appendix, we first construct upper and lower bounds for the minimum Hessian eigenvalue at the TS. In the second part of the Appendix, we introduce an approximation for the eigenvector associated with the minimum Hessian eigenvalue and show that its expectation value provides a tighter upper bound for the minimum Hessian eigenvalue. For clarity, throughout this Appendix we focus on symmetric TS. That is, TS where the zero insertion is made at the same layer $l$ for $\beta$ and $\gamma$ components. We fully describe our construction for the case $l\in [2,p+1]$ and only provide the final expression for the remaining eigenvectors.  

\subsection{Bound on the minimum Hessian eigenvalue}
Given $\Gamma_{\min}^{p}$, a local minima of QAOA$_p$, let $\Gamma_{\text{TS}}^{p+1}(l,l)$ be the TS constructed by padding the $l$-th layer of the QAOA$_p$ circuit with zeros. For clarity, we will omit the layer index whenever possible.

We start by applying a change of basis $\mathcal{P}$ that takes the Hessian at $\Gamma_{\text{TS}}^{p+1}$ to the following generic form:
\begin{align}
H_P(\Gamma_{\text{TS}}^{p+1})\mapsto H(\Gamma_{\text{TS}}^{p+1}) &= \mathcal{P}^T H_P(\Gamma_{\text{TS}}^{p+1}) \mathcal{P} \\
&=\left(\begin{array}{cc}
H(\Gamma_{\text{min}}^{p}) & v(l,l)\\
v^{T}(l,l) & h(l,l)
\end{array}\right).\label{eq:GenericHessianTS}
\end{align}
Ref.~\cite{Sack2022} demonstrated that the $2p\times2$ rectangular matrix $v(l,l)$
is constructed by taking the $l-1$-th and the $p+l$-th columns of
$H(\Gamma_{\text{min}}^{p})$. Using this knowledge, we can apply a combination of elementary transformations $R^D_{i, j}$ on the rows of $H(\Gamma^{p+1}_{\mathrm{TS}})$ as follows: $R_{i,j}^{D}(m)$ is the $D \times D$ identity matrix with an additional non-zero entry $m$ in the $(i,j)$ position. Note that when applied on the left to a matrix $A$, the
resulting matrix will have 
$$
[A]_{i,x}\mapsto[A]_{i,x}+m[A]_{j,x}.
$$The inverse of this matrix is simply $(R_{i,j}^{D}(m))^{-1}=R_{i,j}^{D}(-m)$. Using the above definition, we bring the Hessian at the point $\Gamma_{\text{TS}}^{p+1}$
to a block diagonal form:
\begin{align}
H_{\text{block}} & =\mathcal{R}H(\Gamma_{\text{TS}}^{p+1})\mathcal{R}^{T}=\left(\begin{array}{cc}
H(\Gamma_{\text{min}}^{p}) & 0\\
0 & \bar{h}
\end{array}\right) \label{eq:congruence_rel},\\ 
\mathcal{R} & =R_{2(p+1),p+l}^{2(p+1)}(-1)R_{2p+1,l-1}^{2(p+1)}(-1), \label{eq:definition_Cal_R}\\
\bar{h} & =\left(\begin{array}{cc}
0 & \bar{b}\\
\bar{b} & 0
\end{array}\right),
\end{align}
where 
\begin{multline}
    \label{eq:bbar_definition}
     \bar{b} = \partial_{\beta_l} \partial_{\gamma_l} E(\Gamma^{p+1}_{l,l}) - \partial_{\beta_{l-1}} \partial_{\gamma_l} E(\Gamma^{p}_{\min})  \\
    = \langle + | U^\dagger [H_C , U_{\geq l}[H_C, H_B] U^\dagger_{\geq l}] U|+\rangle.
\end{multline}

The transformation defined above subtracts rows $l-1$ and $p+l$ of $H(\Gamma_{\text{min}}^{p})$ to the first and second row of $v^{T}$ respectively. Then, we apply the same operation but on the columns. It is important to note that the eigenvalues of $H_P(\Gamma_{\text{TS}}^{p+1})$ do change under $\mathcal{R}$. This is because the transformation we applied is not
a similarity transformation. To see this, note that by definition $\mathcal{R}^{T}\neq\mathcal{R}^{-1}$. Instead, we showed that $H_{\mathrm{block}}$ and $H(\Gamma^{p+1}_{\mathrm{TS}})$ are congruent. Thus, $H(\Gamma^{p+1}_{\mathrm{TS}})$ and $H_{\mathrm{block}}$ have the same signature, that is, they have the same number of positive, negative and zero eigenvalues. Since $H(\Gamma^p_{\min})$ is positive definite by assumption, it implies that $H(\Gamma^{p+1}_{\mathrm{TS}})$ has either a unique negative eigenvalue or two zero eigenvalues if and only if $\bar{b}=0$.

Moreover, we can further exploit the above congruence relation by using a result from~\citeauthor{ostrowski} which relates the eigenvalues of $H(\Gamma^{p+1}_{\mathrm{TS}})$ and $H_{\mathrm{block}}$ in the following way:
\begin{equation}
    \lambda_i(H_{\mathrm{block}}) \leq  \theta_i \lambda_i(H(\Gamma^{p+1}_{\mathrm{TS}})),
\end{equation}
where $\lambda_{\min}(\mathcal{R}\mathcal{R}^T) \leq \theta_i \leq \lambda_{\max}(\mathcal{R}\mathcal{R}^T)$, and $\lambda_1 \leq \lambda_2 \leq \cdots \leq \lambda_{2p+2}$. Due to the problem structure it is straightforward to compute the spectrum of the matrix $\mathcal{R}\mathcal{R}^T$ which consists of three different eigenvalues: $\frac{1}{2}(3+\sqrt{5}), \frac{1}{2}(3-\sqrt{5})$ and $1$ with multiplicities $2, 2$ and $2p-2$ respectively. As a consequence, we obtain:
\begin{equation}
    \label{eq:FirstEigBound}
    -\frac{3+\sqrt{5}}{2}|\bar{b}|\leq\lambda_{1}(H(\Gamma_{\text{TS}}^{p+1}))\leq-\frac{2}{3+\sqrt{5}}|\bar{b}|,
\end{equation}
where we used that $\lambda_{1}(H_{\text{block}})=-|\bar{b}|$. Eq.~\eqref{eq:FirstEigBound} provides upper and lower bounds on the magnitude of the minimum Hessian eigenvalue at the TS. Below, we introduce an approximation for the eigenvector associated with the minimum Hessian eigenvalue.
\subsection{Eigenvector approximation}
By definition, the minimum eigenvalue of $H(\Gamma_{\text{TS}}^{p+1})$ is given by
\begin{align}
    \label{eq:exact_eigval}
    \lambda_{1}(H(\Gamma_{\text{TS}}^{p+1})) &= \min_{w} \frac{\langle w, H(\Gamma_{\text{TS}}^{p+1}) w\rangle}{\langle w, w \rangle} \notag, \\
    &= \min_{v} \frac{\langle v, \mathcal{R} H(\Gamma_{\text{TS}}^{p+1}) \mathcal{R}^T v\rangle}{\langle v, \mathcal{R}\mathcal{R}^T v \rangle} \notag, \\
    &= \min_{v} \frac{\langle v, H_{\mathrm{block}} v\rangle}{\langle v, \mathcal{R}\mathcal{R}^T v \rangle},
\end{align}
where we used the congruence relationship from Eq.~\eqref{eq:congruence_rel}. For ease of notation, let us denote by $u \in \mathbb{R}^{2p}$ the first $2p$ components of $v\in \mathbb{R}^{2p+2}$, and by $y = (y_1 \; y_2)^T \in \mathbb{R}^2$ the last two. Using this notation, Eq.~\eqref{eq:exact_eigval} can be written as
\begin{align}
    \label{eq:exact_eigval_v2}
    \lambda_{1}(H(\Gamma_{\text{TS}}^{p+1})) &= \min_{u, y} \frac{\langle u, H(\Gamma_{\min}^{p}) u\rangle + 2\bar{b} y_1 y_2}{1+y_1 (y_1-2u_{l-1})+y_2 (y_2-2u_{p+l})}.
\end{align}
\subsubsection{A simple approximation}
Solving Eq.~\eqref{eq:exact_eigval_v2} requires knowledge of the spectrum of $H(\Gamma^p_{\min})$ which is generally hard to implement in an actual quantum device. Thus, we first aim to provide estimates that do not require access to such information. We start with 
\begin{eqnarray}
    \label{eq:vfirst_approx}
    \mathtt{v}_1^{(0)} = \left( 
    \begin{array}{c}
         0  _{2p}  \\
         1/\sqrt{2} \\
         -\sign(\bar{b})/\sqrt{2}
    \end{array}
    \right),
\end{eqnarray}
which corresponds to the eigenvector of $H_{\mathrm{block}}$ with negative eigenvalue $-|\bar{b}|$. Using this ansatz in Eq.~\eqref{eq:exact_eigval_v2} we obtain $\lambda_{\mathrm{TS}}=-|\bar{b}|/2$ which further tightens the upper bound in Eq.~\eqref{eq:FirstEigBound}. Equivalently, we can obtain an estimate of the exact Hessian eigenvector with Rayleigh coefficient $\lambda_{\mathrm{TS}}$ by transforming $v_{\mathrm{bound}}$ to the original basis, that is, we take $\delta_{\mathrm{TS}}=\mathcal{P}^T\mathcal{R}^T \mathtt{v}_1^{(0)}$ as our estimate for the Hessian eigenvector associated with the unique negative eigenvalue. The specific form of $\delta_{\mathrm{TS}}$ depends on the transition state in consideration and reads:
\begin{equation}
    [\delta_{\rm{TS}}]_j = \begin{cases}
        \frac{1}{2} & \text{if } j=l-1, l\\
        - \frac{\sign(\bar{b})}{2} & \text{if } j = p+l+1, \\
        \frac{\sign(\bar{b})}{2} & \text{if } j = p+l+2, \\
        0 & \text{otherwise}.
        \end{cases}
\end{equation}
Thus, the approximate Hessian eigenvector has weights at layer $l$ and the two adjacent gates. The sparse structure of $\delta_{\rm{TS}}$ comes from the fact that $\mathcal{R}^{T}$ has only two non-zero off-diagonal matrix elements in positions $(2(p+1), p+l), (2p+1, l-1)$. Thus, when acting $\mathtt{v}_1^{(0)}$, we will get a vector with also four non-zero elements. Since $\mathcal{R}^{T}$ is not an orthogonal matrix, then we need to normalize the resulting vector $\mathcal{R}^{T} \mathtt{v}_1^{(0)}$. Finally, the permutation action will reorder the elements in the vector without changing its content. 

Finally, we list without derivation eigenvector and eigenvalue approximations for the remaining TS:
\begin{enumerate}
    \item For $l=1$ we have that:
        \begin{equation}
            [\delta_{\rm{TS}}]_j = 
            \begin{cases}
        -\frac{\sign(b)}{\sqrt{2}} & \text{if } j=l, \\
        \frac{1}{2} & \text{if } j=p+1+l,\\
        -\frac{1}{2} & \text{if } j=p+1+l+1, \\
        0 & \text{otherwise}.
        \end{cases}
        \label{eq:eigvec_bound_ts_1}
        \end{equation}
    with $b=\langle +| [H_C, [H_B, U(\Gamma^p_{\min})^\dagger H_C U(\Gamma^p_{\min})]]|+\rangle$. The approximate eigenvalue in this case equals $-\abs{b}/\sqrt{2}$.

    \item For $l=p+1$ we have that:
        \begin{equation}
            [\delta_{\rm{TS}}]_j = \begin{cases}
        \frac{\sign(b)}{2} & \text{if } j=l-1, \\
        -\frac{\sign(b)}{2} & \text{if } j=l,\\
        \frac{1}{\sqrt{2}} & \text{if } j=p+1+l, \\
        0 & \text{otherwise}.
        \end{cases}
        \end{equation}
    with $b=\langle +| U(\Gamma^p_{\min})^\dagger| [H_C, [H_B, H_C]] U(\Gamma^p_{\min})|+\rangle$. The approximate eigenvalue in this case equals $-\abs{b}/\sqrt{2}$.
\end{enumerate}
We emphasize that the TS with $l=1$ and $l=p+1$ where the identity gates are inserted at the edges of the optimized QAOA circuit are special since the bound has a prefactor $1/\sqrt{2}$ in contrast to the remaining TS where the bound features a prefactor $1/2$. We also emphasize this using a constant $b$ rather than $\bar b$ for the ``bulk'' transition states. 

\subsubsection{Quality of the eigenvector and eigenvalue estimates}
So far we have derived simple estimates for the eigenvector and corresponding negative eigenvalue of the Hessian at each transition state. As we show in the main text, these estimates are generally accurate as measured by the relative error in the eigenvalue and the `infidelity' between the exact and approximate eigenvector. However, there are cases where the accuracy of our estimates drops leading to relative errors of the order of $\sim 10 \% - 30\%$ and eigenvector infidelity becomes $\sim 0.8$. To better understand the source of errors let us go back to Eq.~\eqref{eq:exact_eigval_v2} and expand $u$ as a linear combination of the eigenvectors of $H(\Gamma^p_{\min})$ with eigenvalues $0 < \kappa_1 \leq \cdots \leq \kappa_{2p}$.
\begin{equation}
    \label{eq:eigbasis_hmin}
    u = \sum_{l=1}^{2p} a_l \mathtt{x}_l.
\end{equation}
Using this, we rewrite Eq.~\eqref{eq:exact_eigval_v2} as follows
\begin{align}
    % \label{eq:exact_eigval_v2}
    \lambda_{1}(H(\Gamma_{\text{TS}}^{p+1})) &= \min_{u, y} \frac{\sum_{l=1}^{2p} a^2_l \kappa_l + 2\bar{b} y_1 y_2}{1+y_1 (y_1-2u_{l-1})+y_2 (y_2-2u_{p+l})}.
\end{align}
To simplify things, we assume that $y_1 = \frac{a_0}{\sqrt{2}}$, and $y_2 = -\sign(\bar{b})\frac{a_0}{2}$, based on Eq.~\eqref{eq:vfirst_approx}. Furthermore, we generally expect that the expansion in the eigenbasis of $H(\Gamma^p_{\min})$ is concentrated around the first $k$-th eigenvectors, where the value of $k$ depends on the magnitude of $\bar{b}$. Specifically, we expect $k$ to be such that $\kappa_k \sim \bar{b}$. Moreover, eigenvectors with eigenvalues $\kappa_l \gg |\bar{b}|$ should contribute negligibly to the optimal solution due to their large positive contribution to the numerator. Under these assumptions, we obtain an approximation to the minimum Hessian eigenvalue: 
\begin{align}
    \label{eq:exact_eigval_v3}
    \lambda_{1}(H(\Gamma_{\text{TS}}^{p+1})) &\approx \min_{u, y} \frac{\sum_{l=1}^{k} a^2_l (\kappa_l + \bar{b}) - |\bar{b}|}{2+\sum_{l=1}^k a_l^2 + \sqrt{2}a_0(u_{p+l}-u_{l-1})}.
\end{align}

When minimizing this objective function subject to the normalization constraint, two distinct regimes emerge based on the relative magnitudes of $|\bar{b}|$ and the ordered eigenvalues ${\kappa_1,...,\kappa_k}$:

\begin{enumerate}
\item \textit{Weak coupling regime} ($|\bar{b}| \ll \kappa_1$): In this case, the optimal solution is achieved with $a_0 = 1$ and all other $a_j = 0$, yielding $\lambda_1 \approx -|\bar{b}|/2$. This occurs because any attempt to reduce the denominator through non-zero $a_j$ values would introduce positive $\kappa_j$ terms in the numerator that dominate the potential gain from denominator reduction.
\item \textit{Strong coupling regime} ($|\bar{b}| \gg \kappa_1$): Here, a more negative eigenvalue becomes achievable through a careful balance of competing effects. Non-zero $a_j$ values are chosen to optimize between:
\begin{itemize}
    \item The positive contribution of $\kappa_j$ terms in the numerator,
    \item The reduction of the denominator through appropriate $u_{p+l}$ and $u_{l-1}$ values,
    \item The normalization constraint on the $a_j$ coefficients.
\end{itemize}
\end{enumerate}
\begin{figure}[t]
    \centering
\includegraphics[width=0.99\columnwidth]{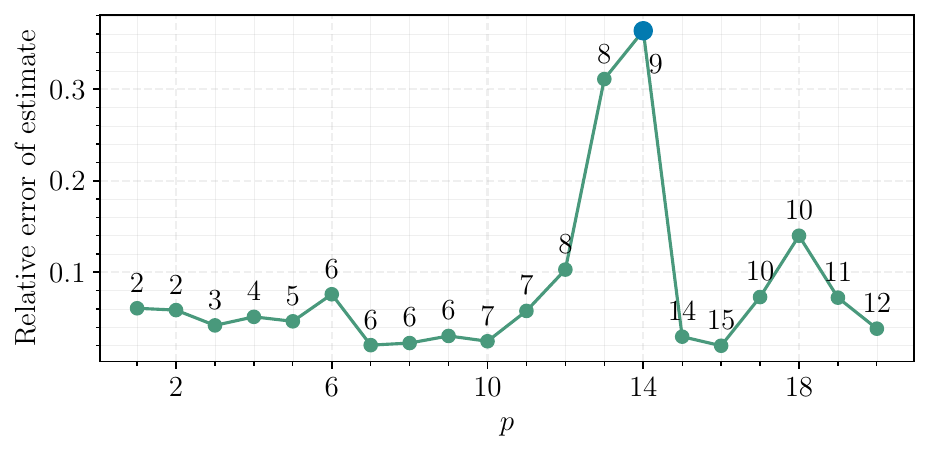}
    \caption{Maximum relative error of the minimum Hessian eigenvalue estimation $-\bar{b}/2$ at transition states with index given by the number on top of the points displayed. The data was obtained from local minima of the QAOA with circuit depths $p\in [1,20]$.}
 \label{fig:rel_error_eigval_instance}
\end{figure}
The transition between these regimes occurs when $|\bar{b}| \approx \kappa_1$, where the denominator reduction begins to compensate for the eigenvalue contributions. In this strong coupling regime, the error in our estimates can become significant ($\sim 10\%-30\%$) due to the complex interplay between multiple eigenvectors.

To showcase such behaviors, we consider an instance of \textsc{MaxCut} on 3-regular graphs with $N=10$ vertices. For each of the local minima obtained through optimization with circuit depths $p \in [1, 20]$, we compute the relative error of our eigenvalue estimate for all transition states ($2p+1$ transition states out of a local minimum at circuit depth $p$). The maximum relative error, together with the indices of the transition states, is displayed in Fig.~\ref{fig:rel_error_eigval_instance}. The relative errors in our eigenvalue estimates typically remain below $10\%$. One notable exception, for example, occurs for the transition state with index $9$ above, which emerges from the QAOA local minimum at circuit depth $p=14$, where we observe a larger error $\sim 36\%$.

\begin{figure}[b]
    \centering
\includegraphics[width=0.99\columnwidth]{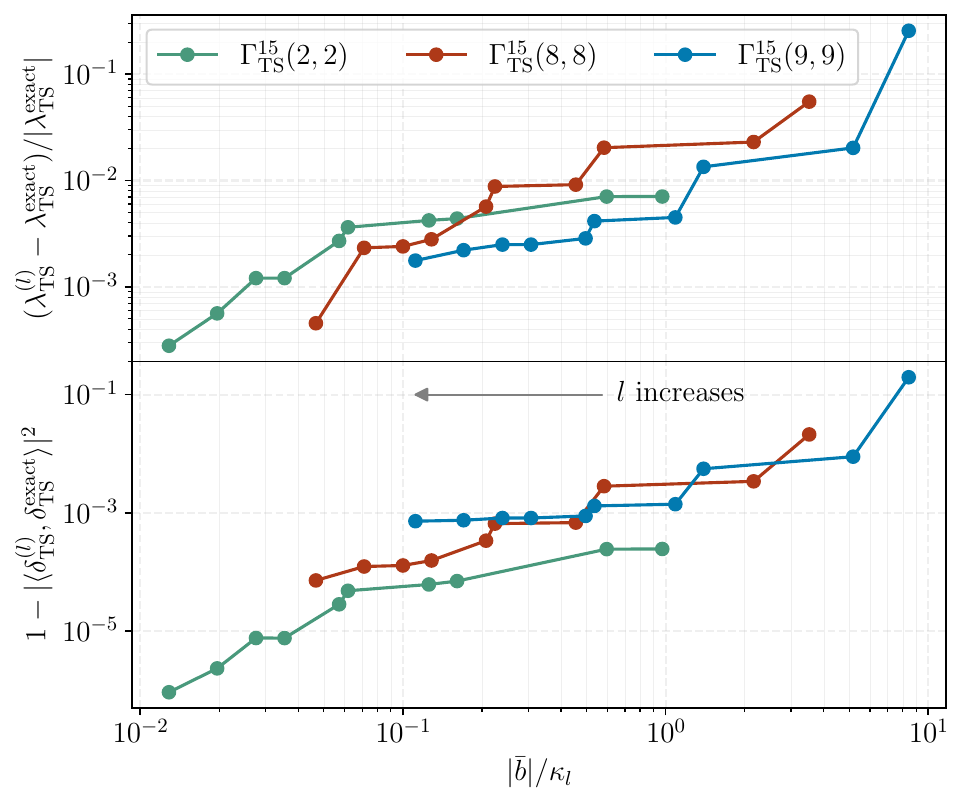}
    \caption{Improvement of the Hessian eigenvalue and eigenvector estimates at a transition state as the number of eigenvectors of the Hessian $H(\Gamma^p_{\min})$ at the corresponding local minimum included in Eq.~\eqref{eq:exact_eigval_v3} increases.}
    \label{fig:estimates_metrics}
\end{figure}

Let us consider the local minimum where our estimate yields the lowest accuracy ($p=14$ in the above figure), and examine transition states with indices $2$, $8$, and $9$ to analyze how accuracy improves as we increase the number of eigenvectors of $H(\Gamma^{14}_{\min})$ in the expansion from Eq.~\eqref{eq:exact_eigval_v3}. As demonstrated in Fig.~\ref{fig:estimates_metrics}, when $\bar{b} \sim \kappa_1$, both eigenvalue and eigenvector estimate errors remain small ($\sim 1\%$). Equivalently, to achieve error estimates of $\sim 1\%$, we must include all eigenvectors of $H(\Gamma^p_{\min})$ satisfying $\kappa_l \lesssim \bar{b}$.
\subsubsection{Effects of eigenvector estimation accuracy on optimization}
\begin{figure}[bt]
    \centering
    \includegraphics[width=\columnwidth]{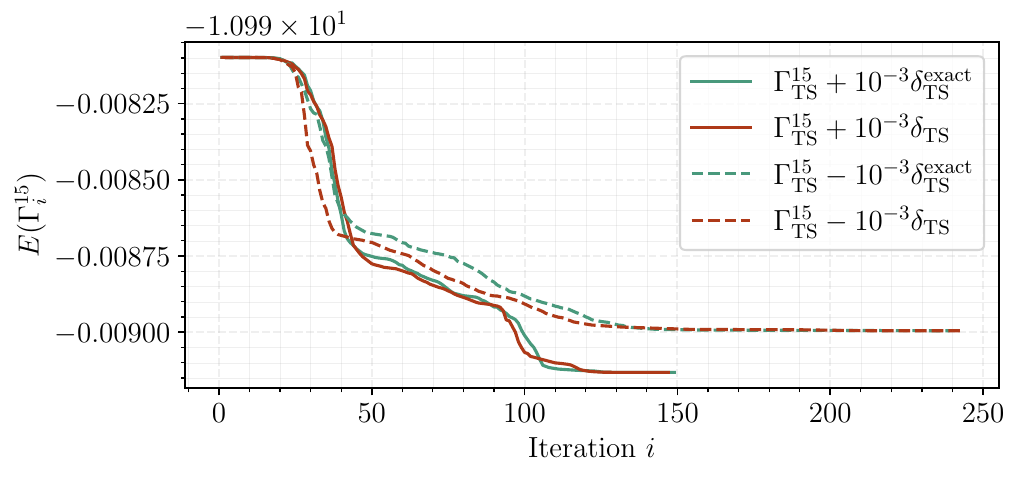}
    \caption{Comparable performance of optimization -as captured by the quality of the converged local minimum, and the number of iterations- when moving a fixed step $\pm 10^{-3}$ along the exact and approximate negative curvature respectively.}
    \label{fig:optim_with_bound}
\end{figure}

Although the eigenvalue and eigenvector estimates presented here typically exhibit high accuracy, certain cases show reduced precision. Identifying such cases would require knowledge of the spectrum of the Hessian at the corresponding local minimum $H(\Gamma^p_{\min})$, which we avoid computing due to the associated measurement overhead. Importantly, we observe no adverse effects on optimization performance even in cases where estimate accuracy decreases significantly. We conjecture that optimization should converge to an equivalent or similar solution provided the overlap with the exact eigenvector remains substantial ($>0.5$). To illustrate this, we examine the optimization from transition state $\Gamma^{15}_{\mathrm{TS}}(9,9)$, where our eigenvalue and eigenvector estimates show errors of $\sim 36\%$ and $\sim 29\%$ respectively. Fig.~\ref{fig:optim_with_bound} demonstrates that optimization from the transition state, using a fixed step along the negative curvature direction, reaches the same local minimum in comparable iterations of the \textsc{BFGS} optimizer: $146$ and $148$ for the exact and approximate Hessian eigenvector approaches, respectively. While we maintain \textsc{BFGS} for consistency with our previous analysis, we note that alternative optimization methods such as \textsc{Adam} exhibit analogous convergence behavior. 

In summary, our results demonstrate that even in scenarios where the accuracy of our estimates decreases, the impact on optimization performance remains negligible at worst. Consequently, we will employ these bounds for the remainder of our analysis, as they provide a practical compromise between computational efficiency and optimization reliability.

\subsubsection{Simplified expression for $b$}
In the next section, we will use the eigenvector approximation Eq.~\eqref{eq:eigvec_bound_ts_1} to derive a lower bound on the energy gain after each iteration of the QAOA algorithm. However, it is essential that before we obtain a simplified expression for $b=\langle +| [H_C, [H_B, U(\Gamma^p_{\min})^\dagger H_C U(\Gamma^p_{\min})]]|+\rangle$. This simplification leverages the specific forms of the mixing Hamiltonian $H_B$ and the cost Hamiltonian $H_C$. We use that both states $|+\rangle$ and $H_C|+\rangle$ are eigenvectors of the mixing Hamiltonian $H_B$ with eigenvalues $-N$ and $-N+4$ respectively. This allows us to simplify the expression for $b$ and arrive at the equation:
\begin{equation}
\label{eq:simplified_b}
    b = 8 \langle +| \tilde{H}_C H_C |+\rangle,
\end{equation}
where $\tilde{H}_C=U(\Gamma^p_{\min})^\dagger H_C U(\Gamma^p_{\min})$ can be thought as the cost Hamiltonian in the Heisenberg picture. Furthermore, the condition $\partial_{\gamma_1} E(\Gamma^{p+1}_{\mathrm{TS}})=2\Im{\langle +| \tilde{H}_C H_C|+\rangle} = 0$, which arises from the transition state being at a stationary point, is applied. It is important to note that the curvature in this scenario is described by $\lambda_{\mathrm{TS}}= -|b|/\sqrt{2}$, equivalently expressed as:
\begin{equation}
\label{eq:curvature_b_final}
    \lambda_{\mathrm{TS}} = -\sign(b) b/\sqrt{2} = -4\sqrt{2}|\langle +| \tilde{H}_C H_C |+\rangle|.
\end{equation}

\section{Expansion of energy alongside the index-1 direction \label{Sec:App-expand}}
Given a local minimum of QAOA$_p$, the transition states construction introduced in~\cite{Sack2022} guarantees that the energy has to decrease alongside the index-1 direction. Thus, in this section, we will use the approximate Hessian eigenvector introduced in Eq.~\eqref{eq:eigvec_bound_ts_1} to provide a lower bound on the energy decrease after optimization of QAOA$_{p+1}$. Out of all possible $2p+1$ transition states available for a given local minima, we focus on the transition state constructed by inserting the zeros in the first layer of the QAOA circuit. 
\subsection{Energy}
We begin by simplifying the expression for the QAOA$_{p+1}$ wave function obtained once we deviate from the transition state along the descent direction:
\begin{multline}
    |\Gamma^{p+1}_{\text{TS}}+\varepsilon \bm{\delta}_{\text{TS}}\rangle  =U(\Gamma^{p+1}_{\text{TS}}+\varepsilon \bm{\delta}_{\text{TS}})|+\rangle \\
    = \Big(\prod_{l=1}^{p+1} U_{B}(\beta_l ) U_C(\gamma_l)\Big) |+\rangle = U(\Gamma^p_{\min}) U_\varepsilon|+\rangle,
\end{multline}
where
\begin{equation}
   U_\varepsilon =   e^{-i \varepsilon/2 H_C} e^{i s_b \sqrt{2}\varepsilon/2 H_B} e^{i \varepsilon/2 H_C},
   \label{eq:u_epsilon}
\end{equation}
and we introduce a short-hand notation:
\begin{equation}
   s_b= \sign(b)
   \label{eq:sign}.
\end{equation}

The next step is to Taylor expand Eq.~\eqref{eq:u_epsilon} around $\varepsilon=0$. For this, we will make use of the following identity:
\begin{align}
    e^{A} B e^{-A} &= B + [A, B] + \frac{1}{2}[A, [A, B]] + \cdots \\
    &+ \frac{1}{n!}[A, [A, \cdots [A, B]\cdots]].
\end{align}
Truncating the above expansion up to the 2nd order, and setting $B=e^{i s_b \varepsilon \sqrt{2} H_B/2}$ and $A=-i \varepsilon H_C/2$ leads to 
\begin{align}
     U_\varepsilon|+\rangle &= e^{-i \varepsilon \phi}|+\rangle - i\frac{\varepsilon}{2}[H_C, e^{i s_b \varepsilon \sqrt{2} H_B/2}]|+\rangle \\
     & - \frac{\varepsilon^2}{2^3}[H_C, [H_C, e^{i s_b \varepsilon \sqrt{2} H_B/2}]]|+\rangle,
\end{align}
where $\phi = s_b \sqrt{2} N/2$ and we used that $H_B|+\rangle = -N|+\rangle$.

To simplify the above expression we need to understand the action of the operator $e^{i \varepsilon \sign(b) \sqrt{2} H_B /2}$ on $H_C$ and $H_C^2$ operators. For this, it is important to note that $H_C|+\rangle$ is an eigenvector of $H_B$ with eigenvalue $(-N+4)$. With this at hand, we obtain that:
\begin{multline}
    \label{eq:uepsilon_1}
     e^{i s_b \varepsilon \sqrt{2} H_B /2} H_C |+\rangle = e^{i s_b \varepsilon (-N+4)\sqrt{2}/2}H_C |+\rangle \\
     = e^{-i \varepsilon \phi} e^{i \varepsilon s_b 2\sqrt{2}}H_C |+\rangle.
\end{multline}
The procedure is slightly more involved in the case of $H_C^2$. This is because $H_C^2$ is a sum of $4$-local, $2$-local, and $0$-local (constant) Hamiltonian densities. More specifically, the squared cost function Hamiltonian is written as
\begin{equation}
    H_C^2 = n_C^2 \mathbb{I} + T_2 + T_4,
\end{equation}
where $T_k$ with $k=2,4$ is a sum involving $4 n_{\mathcal{E}}(\mathcal{G})$ and $n_{\mathcal{E}}(\mathcal{G})(n_{\mathcal{E}}(\mathcal{G}) - 5)$ $k$-local terms respectively. This, together with the $n_{\mathcal{E}}(\mathcal{G})$ terms contributing to $n_C^2 \mathbb{I}$ makes for a total of $n_{\mathcal{E}}(\mathcal{G})^2$ terms. Thus, we obtain
\begin{multline}
    \label{eq:uepsilon_2}
    e^{i s_b \varepsilon \sqrt{2} H_B /2} H_C^2 |+\rangle 
    \\
    =e^{-i \varepsilon \phi}(T_0 \mathbb{I} + e^{i \varepsilon s_b 2 \sqrt{2}}T_2 + e^{i \varepsilon s_b 4 \sqrt{2}} T_4)|+\rangle
    \\
    =e^{-i \varepsilon \phi} (H_C^2 + (e^{i \varepsilon s_b 2 \sqrt{2}}-1) T_2 + (e^{i \varepsilon s_b 4 \sqrt{2}}-1) T_4)|+\rangle \\
    = e^{-i \varepsilon \phi} H_C^2 |+\rangle + e^{-i \varepsilon \phi} O_\varepsilon |+\rangle,
\end{multline}
where we defined 
$$
O_\varepsilon =(e^{i \varepsilon s_b 2 \sqrt{2}}-1) T_2 + (e^{i \varepsilon s_b 4 \sqrt{2}}-1) T_4).
$$
Analogously, we obtain:
\begin{multline}
\label{eq:uepsilon_3}
    [H_C, [H_C, e^{-i \varepsilon s_b\sqrt{2} H_B/2 }]]|+\rangle 
    \\
    = 2e^{-i \varepsilon \phi}( 1-e^{i \varepsilon s_b 2\sqrt{2}})H_C^2|+\rangle + e^{-i\varepsilon \phi} O_\varepsilon |+\rangle.
\end{multline}
Putting together Eq.~\eqref{eq:uepsilon_1}, Eq.~\eqref{eq:uepsilon_2}, and Eq.~\eqref{eq:uepsilon_3} we have a final expression for $U_\varepsilon|+\rangle$ that (up to a global phase) reads
\begin{multline}
\label{eq:u_epsilon_final}
    U_\varepsilon |+\rangle =|+\rangle - i \frac{\varepsilon}{2}( 1-e^{i \varepsilon  s_b 2\sqrt{2}}) H_C |+\rangle 
    \\ - \frac{\varepsilon^2}{2^2}( 1-e^{i \varepsilon s_b 2\sqrt{2}})H_C^2|+\rangle 
     - \frac{\varepsilon^2}{2^3}O_\varepsilon |+\rangle.
\end{multline}
We can then use Eq.~\eqref{eq:u_epsilon_final} to obtain the expression for the energy when moving along the index-1 direction as a function of $\varepsilon$:
\begin{multline}
        E(\Gamma^{p+1}_{\mathrm{TS}} + \varepsilon \delta_{\mathrm{TS}}) 
        =\langle +|U_\varepsilon^\dagger U^\dagger(\Gamma^{p}_{\min}) H_C U(\Gamma^{p}_{\min}) U_\varepsilon |+\rangle
        \\
        = E(\Gamma^{p}_{\min}) - \varepsilon \sin(2\sqrt{2} s_b\varepsilon) \langle +| \tilde{H}_C H_C|+\rangle 
        \\
        + \frac{\varepsilon^2}{2} \sin(\varepsilon s_b \sqrt{2})^2 \partial^2_{\gamma_1} E(\Gamma^{p+1}_{\rm{TS}}) 
        \\
        - \frac{\varepsilon^2}{2}\sin(\varepsilon 2 \sqrt{2} s_b)\Im{\langle +|\tilde{H}_C H_C^2|+\rangle} 
        \\
        - \frac{\varepsilon^2}{4} \Re{\langle +| \tilde{H}_C O_\varepsilon |+\rangle},
\end{multline}
where for ease of notation we introduced $\tilde{O} = U^\dagger(\Gamma) O U(\Gamma)$ for a generic Hermitian operator $O$. The next step in the calculation is to isolate terms depending on the magnitude of their contribution with circuit depth $p$ (independently of the value of $\varepsilon$). For this, we need to further simplify the expectation value $\langle +|\tilde{H}_C O_\varepsilon|+\rangle$. After performing careful algebraic manipulations, we obtain a convoluted expression for the energy along the index-1 direction, represented as the fourth-order polynomial in the parameter $\varepsilon$. For enhanced readability, we present the terms of the expansion of $\Delta E(\varepsilon) = E(\Gamma^{p+1}_{\mathrm{TS}} + \varepsilon \delta_{\mathrm{TS}}) - E(\Gamma^p_{\min})$ separately, organized according to the power of $\varepsilon$ with which they are associated:
\begin{widetext}
\begin{align}
    \varepsilon^2 \to &-\varepsilon \sin(2 \sqrt{2} s_b\varepsilon) \langle +| \tilde{H}_C H_C|+\rangle \\
    \varepsilon^3 \to &-\frac{\varepsilon^2}{4}\sin(2 \sqrt{2} s_b\varepsilon) \Im{\langle +|\tilde{H}_C T_2 |+\rangle}
    -\frac{\varepsilon^2}{2}\sin(2 \sqrt{2}s_b \varepsilon)\Big(1-\frac{\sin(4\sqrt{2}s_b \varepsilon)}{2\sin(2\sqrt{2} s_b\varepsilon)}\Big)\Im{\langle +|\tilde{H}_C T_4 |+\rangle}. \\
    \varepsilon^4 \to &\frac{\varepsilon^2}{2} \sin^2(\varepsilon s_b\sqrt{2}) \partial^2_{\gamma_1} E(\Gamma^{p+1}_{\rm{TS}}) + \frac{\varepsilon^2}{4} 2\sin^2(\varepsilon s_b\sqrt{2}) \Re{\langle +|\tilde{H}_C T_2|+\rangle}  +\frac{\varepsilon^2}{2}\sin^2(2\varepsilon s_b\sqrt{2}) \Re{\langle +|\tilde{H}_C T_4|+\rangle} . \label{eq:c_terms}
\end{align}    
\end{widetext}
From the above equations, we see that at the first non-trivial order in $\varepsilon$ the only cubical $\sim\varepsilon^3$ contribution that remains is $\Im{\langle +| \tilde{H}_C T_2|+\rangle}$. The quartic $\sim\varepsilon^4$ term is however more involved, and we resort instead to numerically verifying the order of magnitude of each term involved as a function of circuit depth $p$. 

\begin{figure}[t]
    \centering
    \includegraphics[width=0.99\linewidth]{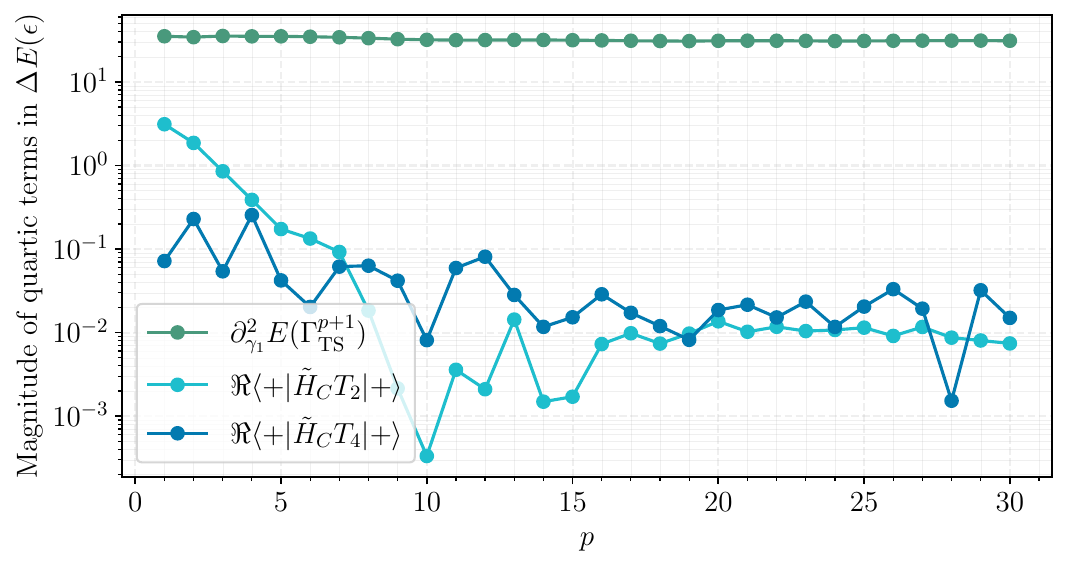}
    \caption{Magnitude of the prefactors of three different quartic terms $\sim \varepsilon^4$ in the energy expansion along the index-1 direction as a function of the circuit depth $p$. The first term in the expansion is dominant.}
    \label{fig:quartic_terms}
\end{figure}

In Figure~\ref{fig:quartic_terms} we show the circuit depth dependence of three different terms, (1) $\partial^2_{\gamma_1} E(\Gamma^{p+1}_{\rm{TS}})$, (2) $\Re{\langle +|\tilde{H}_C T_2|+\rangle}$ and (3) $\Re{\langle +|\tilde{H}_C T_4|+\rangle}$ for a single \textsc{MaxCut} instance of a 3-regular weighted graph with $N=14$ vertices (see Fig.~\ref{fig:harvard_instance} for the details of the instance). The numerical data reveals that the term (1) $\partial^2_{\gamma_1} E(\Gamma^{p+1}_{\rm{TS}})$ dominates over terms (2)-(3)  at all circuit depths. 

Finally,  we obtain a close and concise expression for the energy change along the index-1 direction:
\begin{align}
\label{eq:energy_slice_full}
    \Delta E(\varepsilon) &\approx -\varepsilon \sin(2 \sqrt{2} s_b\varepsilon) \langle +| \tilde{H}_C H_C|+\rangle \notag \\
    &-\frac{\varepsilon^2}{4}\sin(2 \sqrt{2}s_b \varepsilon) \Im{\langle +|\tilde{H}_C T_2 |+\rangle} \notag\\
    &+ \frac{\varepsilon^2}{2} \sin^2(\varepsilon s_b \sqrt{2}) \partial^2_{\gamma_1} E(\Gamma^{p+1}_{\rm{TS}}).
\end{align}
It is worth noting that in the above equation, Eq.~\eqref{eq:energy_slice_full} the quadratic term at small $\varepsilon$ is negative and proportional to the (approximate) minimum curvature, i.e,
\begin{equation}
    \varepsilon^2 \to - \varepsilon^2 2\sqrt{2} \sign(b) b = \varepsilon^2 \frac{\lambda_{\mathrm{TS}}}{2}.   
\end{equation}
The negative value of the second order term provided that the fourth order term is positive (see discussion in Sec.~\ref{sec:iv_B} after Eq.~\eqref{eq:c_approx}), leads to an existence of non-trivial minimum in the expansion of the energy along the index-1 direction. Our argument as to why $\partial^2_{\gamma_1} E(\Gamma^{p+1}_{\rm{TS}}) > 0$ lies on the observation that it can be approximated as the positive energy difference (see Eq.~\eqref{eq:c_approx}) of the states $\frac{1}{n_C}U H_C|+\rangle$ and $U|+\rangle$.

\begin{figure}[b]
    \centering
    \includegraphics[width=0.99\linewidth]{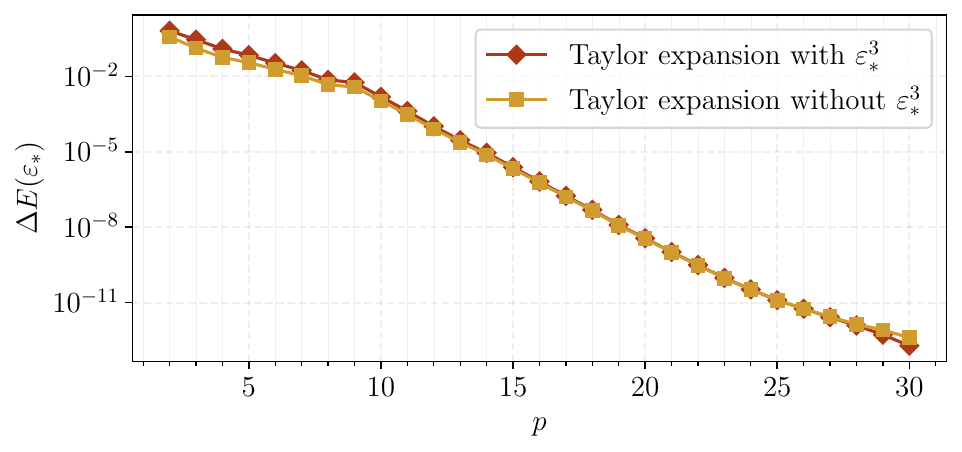}
    \caption{The energy difference between the transition state and the local minima obtained along the descent direction shows little sensitivity to the presence of the cubic term in the expansion. 
    }
    \label{fig:effect_of_cubic_term}
\end{figure}

As discussed in the main text, we discard the cubic $\sim \varepsilon^3$ term for simplicity.  This approximation is justified by the negligible effect of the cubic term in the minima of the energy along the index-1 direction, as shown in Fig.~\ref{fig:effect_of_cubic_term}. We are then left with an expression for the energy $\Delta E(\varepsilon)_{\rm{sym}}$ which has two degenerate global minima
\begin{align}
    \Delta E(\varepsilon^*)_{\rm{sym}}&= -\frac{\lambda_{\mathrm{TS}}^2}{16 \partial^2_{\gamma_1}E(\Gamma^{p+1}_{\mathrm{TS}})},
    \label{eq:energ_gain_ts_1}
\end{align}
where $(\varepsilon^*)^2 = -\lambda_{\mathrm{TS}}/4\partial^2_{\gamma_1} (\Gamma^{p+1}_{\mathrm{TS}})$.

\bibliography{QAOA-improvementNotes}

\begin{thebibliography}{41}%
\makeatletter
\providecommand \@ifxundefined [1]{%
 \@ifx{#1\undefined}
}%
\providecommand \@ifnum [1]{%
 \ifnum #1\expandafter \@firstoftwo
 \else \expandafter \@secondoftwo
 \fi
}%
\providecommand \@ifx [1]{%
 \ifx #1\expandafter \@firstoftwo
 \else \expandafter \@secondoftwo
 \fi
}%
\providecommand \natexlab [1]{#1}%
\providecommand \enquote  [1]{``#1''}%
\providecommand \bibnamefont  [1]{#1}%
\providecommand \bibfnamefont [1]{#1}%
\providecommand \citenamefont [1]{#1}%
\providecommand \href@noop [0]{\@secondoftwo}%
\providecommand \href [0]{\begingroup \@sanitize@url \@href}%
\providecommand \@href[1]{\@@startlink{#1}\@@href}%
\providecommand \@@href[1]{\endgroup#1\@@endlink}%
\providecommand \@sanitize@url [0]{\catcode `\\12\catcode `\$12\catcode
  `\&12\catcode `\#12\catcode `\^12\catcode `\_12\catcode `\%12\relax}%
\providecommand \@@startlink[1]{}%
\providecommand \@@endlink[0]{}%
\providecommand \url  [0]{\begingroup\@sanitize@url \@url }%
\providecommand \@url [1]{\endgroup\@href {#1}{\urlprefix }}%
\providecommand \urlprefix  [0]{URL }%
\providecommand \Eprint [0]{\href }%
\providecommand \doibase [0]{https://doi.org/}%
\providecommand \selectlanguage [0]{\@gobble}%
\providecommand \bibinfo  [0]{\@secondoftwo}%
\providecommand \bibfield  [0]{\@secondoftwo}%
\providecommand \translation [1]{[#1]}%
\providecommand \BibitemOpen [0]{}%
\providecommand \bibitemStop [0]{}%
\providecommand \bibitemNoStop [0]{.\EOS\space}%
\providecommand \EOS [0]{\spacefactor3000\relax}%
\providecommand \BibitemShut  [1]{\csname bibitem#1\endcsname}%
\let\auto@bib@innerbib\@empty
%</preamble>
\bibitem [{\citenamefont {Cerezo}\ \emph {et~al.}(2021)\citenamefont {Cerezo},
  \citenamefont {Arrasmith}, \citenamefont {Babbush}, \citenamefont {Benjamin},
  \citenamefont {Endo}, \citenamefont {Fujii}, \citenamefont {McClean},
  \citenamefont {Mitarai}, \citenamefont {Yuan}, \citenamefont {Cincio},\ and\
  \citenamefont {Coles}}]{Cerezo_2021}%
  \BibitemOpen
  \bibfield  {author} {\bibinfo {author} {\bibfnamefont {M.}~\bibnamefont
  {Cerezo}}, \bibinfo {author} {\bibfnamefont {A.}~\bibnamefont {Arrasmith}},
  \bibinfo {author} {\bibfnamefont {R.}~\bibnamefont {Babbush}}, \bibinfo
  {author} {\bibfnamefont {S.~C.}\ \bibnamefont {Benjamin}}, \bibinfo {author}
  {\bibfnamefont {S.}~\bibnamefont {Endo}}, \bibinfo {author} {\bibfnamefont
  {K.}~\bibnamefont {Fujii}}, \bibinfo {author} {\bibfnamefont {J.~R.}\
  \bibnamefont {McClean}}, \bibinfo {author} {\bibfnamefont {K.}~\bibnamefont
  {Mitarai}}, \bibinfo {author} {\bibfnamefont {X.}~\bibnamefont {Yuan}},
  \bibinfo {author} {\bibfnamefont {L.}~\bibnamefont {Cincio}},\ and\ \bibinfo
  {author} {\bibfnamefont {P.~J.}\ \bibnamefont {Coles}},\ }\bibfield  {title}
  {\bibinfo {title} {Variational quantum algorithms},\ }\href
  {https://doi.org/10.1038/s42254-021-00348-9} {\bibfield  {journal} {\bibinfo
  {journal} {Nature Reviews Physics}\ }\textbf {\bibinfo {volume} {3}},\
  \bibinfo {pages} {625–644} (\bibinfo {year} {2021})}\BibitemShut {NoStop}%
\bibitem [{\citenamefont {Bharti}\ \emph {et~al.}(2022)\citenamefont {Bharti},
  \citenamefont {Cervera-Lierta}, \citenamefont {Kyaw}, \citenamefont {Haug},
  \citenamefont {Alperin-Lea}, \citenamefont {Anand}, \citenamefont {Degroote},
  \citenamefont {Heimonen}, \citenamefont {Kottmann}, \citenamefont {Menke},
  \citenamefont {Mok}, \citenamefont {Sim}, \citenamefont {Kwek},\ and\
  \citenamefont {Aspuru-Guzik}}]{Bharti_2022}%
  \BibitemOpen
  \bibfield  {author} {\bibinfo {author} {\bibfnamefont {K.}~\bibnamefont
  {Bharti}}, \bibinfo {author} {\bibfnamefont {A.}~\bibnamefont
  {Cervera-Lierta}}, \bibinfo {author} {\bibfnamefont {T.~H.}\ \bibnamefont
  {Kyaw}}, \bibinfo {author} {\bibfnamefont {T.}~\bibnamefont {Haug}}, \bibinfo
  {author} {\bibfnamefont {S.}~\bibnamefont {Alperin-Lea}}, \bibinfo {author}
  {\bibfnamefont {A.}~\bibnamefont {Anand}}, \bibinfo {author} {\bibfnamefont
  {M.}~\bibnamefont {Degroote}}, \bibinfo {author} {\bibfnamefont
  {H.}~\bibnamefont {Heimonen}}, \bibinfo {author} {\bibfnamefont {J.~S.}\
  \bibnamefont {Kottmann}}, \bibinfo {author} {\bibfnamefont {T.}~\bibnamefont
  {Menke}}, \bibinfo {author} {\bibfnamefont {W.-K.}\ \bibnamefont {Mok}},
  \bibinfo {author} {\bibfnamefont {S.}~\bibnamefont {Sim}}, \bibinfo {author}
  {\bibfnamefont {L.-C.}\ \bibnamefont {Kwek}},\ and\ \bibinfo {author}
  {\bibfnamefont {A.}~\bibnamefont {Aspuru-Guzik}},\ }\bibfield  {title}
  {\bibinfo {title} {Noisy intermediate-scale quantum algorithms},\ }\href
  {https://doi.org/10.1103/RevModPhys.94.015004} {\bibfield  {journal}
  {\bibinfo  {journal} {Rev. Mod. Phys.}\ }\textbf {\bibinfo {volume} {94}},\
  \bibinfo {pages} {015004} (\bibinfo {year} {2022})}\BibitemShut {NoStop}%
\bibitem [{\citenamefont {{Preskill}}(2018)}]{preskill2018quantum}%
  \BibitemOpen
  \bibfield  {author} {\bibinfo {author} {\bibfnamefont {J.}~\bibnamefont
  {{Preskill}}},\ }\bibfield  {title} {\bibinfo {title} {{Quantum Computing in
  the NISQ era and beyond}},\ }\href@noop {} {\bibfield  {journal} {\bibinfo
  {journal} {arXiv e-prints}\ ,\ \bibinfo {eid} {arXiv:1801.00862}} (\bibinfo
  {year} {2018})},\ \Eprint {https://arxiv.org/abs/1801.00862}
  {arXiv:1801.00862 [quant-ph]} \BibitemShut {NoStop}%
\bibitem [{\citenamefont {{Farhi}}\ \emph {et~al.}(2014)\citenamefont
  {{Farhi}}, \citenamefont {{Goldstone}},\ and\ \citenamefont
  {{Gutmann}}}]{farhi2014quantum}%
  \BibitemOpen
  \bibfield  {author} {\bibinfo {author} {\bibfnamefont {E.}~\bibnamefont
  {{Farhi}}}, \bibinfo {author} {\bibfnamefont {J.}~\bibnamefont
  {{Goldstone}}},\ and\ \bibinfo {author} {\bibfnamefont {S.}~\bibnamefont
  {{Gutmann}}},\ }\bibfield  {title} {\bibinfo {title} {{A Quantum Approximate
  Optimization Algorithm}},\ }\href@noop {} {\bibfield  {journal} {\bibinfo
  {journal} {arXiv e-prints}\ ,\ \bibinfo {eid} {arXiv:1411.4028}} (\bibinfo
  {year} {2014})},\ \Eprint {https://arxiv.org/abs/1411.4028} {arXiv:1411.4028
  [quant-ph]} \BibitemShut {NoStop}%
\bibitem [{\citenamefont {{Peruzzo}}\ \emph {et~al.}(2014)\citenamefont
  {{Peruzzo}}, \citenamefont {{McClean}}, \citenamefont {{Shadbolt}},
  \citenamefont {{Yung}}, \citenamefont {{Zhou}}, \citenamefont {{Love}},
  \citenamefont {{Aspuru-Guzik}},\ and\ \citenamefont
  {{O'Brien}}}]{peruzzo2014vqe}%
  \BibitemOpen
  \bibfield  {author} {\bibinfo {author} {\bibfnamefont {A.}~\bibnamefont
  {{Peruzzo}}}, \bibinfo {author} {\bibfnamefont {J.}~\bibnamefont
  {{McClean}}}, \bibinfo {author} {\bibfnamefont {P.}~\bibnamefont
  {{Shadbolt}}}, \bibinfo {author} {\bibfnamefont {M.-H.}\ \bibnamefont
  {{Yung}}}, \bibinfo {author} {\bibfnamefont {X.-Q.}\ \bibnamefont {{Zhou}}},
  \bibinfo {author} {\bibfnamefont {P.~J.}\ \bibnamefont {{Love}}}, \bibinfo
  {author} {\bibfnamefont {A.}~\bibnamefont {{Aspuru-Guzik}}},\ and\ \bibinfo
  {author} {\bibfnamefont {J.~L.}\ \bibnamefont {{O'Brien}}},\ }\bibfield
  {title} {\bibinfo {title} {{A variational eigenvalue solver on a photonic
  quantum processor}},\ }\href {https://doi.org/10.1038/ncomms5213} {\bibfield
  {journal} {\bibinfo  {journal} {Nature Communications}\ }\textbf {\bibinfo
  {volume} {5}},\ \bibinfo {eid} {4213} (\bibinfo {year} {2014})},\ \Eprint
  {https://arxiv.org/abs/1304.3061} {arXiv:1304.3061 [quant-ph]} \BibitemShut
  {NoStop}%
\bibitem [{\citenamefont {Farhi}\ \emph {et~al.}(2015)\citenamefont {Farhi},
  \citenamefont {Goldstone},\ and\ \citenamefont {Gutmann}}]{farhi2015quantum}%
  \BibitemOpen
  \bibfield  {author} {\bibinfo {author} {\bibfnamefont {E.}~\bibnamefont
  {Farhi}}, \bibinfo {author} {\bibfnamefont {J.}~\bibnamefont {Goldstone}},\
  and\ \bibinfo {author} {\bibfnamefont {S.}~\bibnamefont {Gutmann}},\
  }\href@noop {} {\bibinfo {title} {A quantum approximate optimization
  algorithm applied to a bounded occurrence constraint problem}} (\bibinfo
  {year} {2015}),\ \Eprint {https://arxiv.org/abs/1412.6062} {arXiv:1412.6062
  [quant-ph]} \BibitemShut {NoStop}%
\bibitem [{\citenamefont {{Farhi}}\ \emph {et~al.}(2020)\citenamefont
  {{Farhi}}, \citenamefont {{Gamarnik}},\ and\ \citenamefont
  {{Gutmann}}}]{farhi2020wholegraph}%
  \BibitemOpen
  \bibfield  {author} {\bibinfo {author} {\bibfnamefont {E.}~\bibnamefont
  {{Farhi}}}, \bibinfo {author} {\bibfnamefont {D.}~\bibnamefont
  {{Gamarnik}}},\ and\ \bibinfo {author} {\bibfnamefont {S.}~\bibnamefont
  {{Gutmann}}},\ }\bibfield  {title} {\bibinfo {title} {{The Quantum
  Approximate Optimization Algorithm Needs to See the Whole Graph: A Typical
  Case}},\ }\href@noop {} {\bibfield  {journal} {\bibinfo  {journal} {arXiv
  e-prints}\ ,\ \bibinfo {eid} {arXiv:2004.09002}} (\bibinfo {year} {2020})},\
  \Eprint {https://arxiv.org/abs/2004.09002} {arXiv:2004.09002 [quant-ph]}
  \BibitemShut {NoStop}%
\bibitem [{\citenamefont {Boulebnane}\ and\ \citenamefont
  {Montanaro}(2021)}]{boulebnane2021predicting}%
  \BibitemOpen
  \bibfield  {author} {\bibinfo {author} {\bibfnamefont {S.}~\bibnamefont
  {Boulebnane}}\ and\ \bibinfo {author} {\bibfnamefont {A.}~\bibnamefont
  {Montanaro}},\ }\href@noop {} {\bibinfo {title} {Predicting parameters for
  the quantum approximate optimization algorithm for max-cut from the
  infinite-size limit}} (\bibinfo {year} {2021}),\ \Eprint
  {https://arxiv.org/abs/2110.10685} {arXiv:2110.10685 [quant-ph]} \BibitemShut
  {NoStop}%
\bibitem [{\citenamefont {Boulebnane}\ and\ \citenamefont
  {Montanaro}(2022)}]{boulebnane2022solving}%
  \BibitemOpen
  \bibfield  {author} {\bibinfo {author} {\bibfnamefont {S.}~\bibnamefont
  {Boulebnane}}\ and\ \bibinfo {author} {\bibfnamefont {A.}~\bibnamefont
  {Montanaro}},\ }\href@noop {} {\bibinfo {title} {Solving boolean
  satisfiability problems with the quantum approximate optimization algorithm}}
  (\bibinfo {year} {2022}),\ \Eprint {https://arxiv.org/abs/2208.06909}
  {arXiv:2208.06909 [quant-ph]} \BibitemShut {NoStop}%
\bibitem [{\citenamefont {{Brand\~{a}o}}\ \emph {et~al.}(2018)\citenamefont
  {{Brand\~{a}o}}, \citenamefont {{Broughton}}, \citenamefont {{Farhi}},
  \citenamefont {{Gutmann}},\ and\ \citenamefont {{Neven}}}]{brandao2018fixed}%
  \BibitemOpen
  \bibfield  {author} {\bibinfo {author} {\bibfnamefont {F.~G.~S.~L.}\
  \bibnamefont {{Brand\~{a}o}}}, \bibinfo {author} {\bibfnamefont
  {M.}~\bibnamefont {{Broughton}}}, \bibinfo {author} {\bibfnamefont
  {E.}~\bibnamefont {{Farhi}}}, \bibinfo {author} {\bibfnamefont
  {S.}~\bibnamefont {{Gutmann}}},\ and\ \bibinfo {author} {\bibfnamefont
  {H.}~\bibnamefont {{Neven}}},\ }\bibfield  {title} {\bibinfo {title} {{For
  Fixed Control Parameters the Quantum Approximate Optimization Algorithm's
  Objective Function Value Concentrates for Typical Instances}},\ }\href@noop
  {} {\bibfield  {journal} {\bibinfo  {journal} {arXiv e-prints}\ ,\ \bibinfo
  {eid} {arXiv:1812.04170}} (\bibinfo {year} {2018})},\ \Eprint
  {https://arxiv.org/abs/1812.04170} {arXiv:1812.04170 [quant-ph]} \BibitemShut
  {NoStop}%
\bibitem [{\citenamefont {{Wurtz}}\ and\ \citenamefont
  {{Love}}(2021)}]{wurtz2021maxcut}%
  \BibitemOpen
  \bibfield  {author} {\bibinfo {author} {\bibfnamefont {J.}~\bibnamefont
  {{Wurtz}}}\ and\ \bibinfo {author} {\bibfnamefont {P.}~\bibnamefont
  {{Love}}},\ }\bibfield  {title} {\bibinfo {title} {{MaxCut quantum
  approximate optimization algorithm performance guarantees for {$p>1$}}},\
  }\href {https://doi.org/10.1103/PhysRevA.103.042612} {\bibfield  {journal}
  {\bibinfo  {journal} {\pra}\ }\textbf {\bibinfo {volume} {103}},\ \bibinfo
  {eid} {042612} (\bibinfo {year} {2021})},\ \Eprint
  {https://arxiv.org/abs/2010.11209} {arXiv:2010.11209 [quant-ph]} \BibitemShut
  {NoStop}%
\bibitem [{\citenamefont {Zhou}\ \emph {et~al.}(2024)\citenamefont {Zhou},
  \citenamefont {Basso},\ and\ \citenamefont {Mei}}]{zhou2024statistical}%
  \BibitemOpen
  \bibfield  {author} {\bibinfo {author} {\bibfnamefont {L.}~\bibnamefont
  {Zhou}}, \bibinfo {author} {\bibfnamefont {J.}~\bibnamefont {Basso}},\ and\
  \bibinfo {author} {\bibfnamefont {S.}~\bibnamefont {Mei}},\ }\href@noop {}
  {\bibinfo {title} {Statistical estimation in the spiked tensor model via the
  quantum approximate optimization algorithm}} (\bibinfo {year} {2024}),\
  \Eprint {https://arxiv.org/abs/2402.19456} {arXiv:2402.19456 [quant-ph]}
  \BibitemShut {NoStop}%
\bibitem [{\citenamefont {Basso}\ \emph
  {et~al.}(2022{\natexlab{a}})\citenamefont {Basso}, \citenamefont {Gamarnik},
  \citenamefont {Mei},\ and\ \citenamefont {Zhou}}]{Basso_2022}%
  \BibitemOpen
  \bibfield  {author} {\bibinfo {author} {\bibfnamefont {J.}~\bibnamefont
  {Basso}}, \bibinfo {author} {\bibfnamefont {D.}~\bibnamefont {Gamarnik}},
  \bibinfo {author} {\bibfnamefont {S.}~\bibnamefont {Mei}},\ and\ \bibinfo
  {author} {\bibfnamefont {L.}~\bibnamefont {Zhou}},\ }\bibfield  {title}
  {\bibinfo {title} {Performance and limitations of the qaoa at constant levels
  on large sparse hypergraphs and spin glass models},\ }in\ \href
  {https://doi.org/10.1109/FOCS54457.2022.00039} {\emph {\bibinfo {booktitle}
  {2022 IEEE 63rd Annual Symposium on Foundations of Computer Science
  (FOCS)}}}\ (\bibinfo {year} {2022})\ pp.\ \bibinfo {pages}
  {335--343}\BibitemShut {NoStop}%
\bibitem [{\citenamefont {Basso}\ \emph
  {et~al.}(2022{\natexlab{b}})\citenamefont {Basso}, \citenamefont {Farhi},
  \citenamefont {Marwaha}, \citenamefont {Villalonga},\ and\ \citenamefont
  {Zhou}}]{Basso_2022_SKmodel}%
  \BibitemOpen
  \bibfield  {author} {\bibinfo {author} {\bibfnamefont {J.}~\bibnamefont
  {Basso}}, \bibinfo {author} {\bibfnamefont {E.}~\bibnamefont {Farhi}},
  \bibinfo {author} {\bibfnamefont {K.}~\bibnamefont {Marwaha}}, \bibinfo
  {author} {\bibfnamefont {B.}~\bibnamefont {Villalonga}},\ and\ \bibinfo
  {author} {\bibfnamefont {L.}~\bibnamefont {Zhou}},\ }\bibfield  {title}
  {{\selectlanguage {en}\bibinfo {title} {The quantum approximate optimization
  algorithm at high depth for maxcut on large-girth regular graphs and the
  sherrington-kirkpatrick model}}\ }(\bibinfo  {publisher} {Schloss Dagstuhl
  – Leibniz-Zentrum für Informatik},\ \bibinfo {year} {2022})\BibitemShut
  {NoStop}%
\bibitem [{\citenamefont {Sureshbabu}\ \emph {et~al.}(2024)\citenamefont
  {Sureshbabu}, \citenamefont {Herman}, \citenamefont {Shaydulin},
  \citenamefont {Basso}, \citenamefont {Chakrabarti}, \citenamefont {Sun},\
  and\ \citenamefont {Pistoia}}]{Sureshbabu_2024}%
  \BibitemOpen
  \bibfield  {author} {\bibinfo {author} {\bibfnamefont {S.~H.}\ \bibnamefont
  {Sureshbabu}}, \bibinfo {author} {\bibfnamefont {D.}~\bibnamefont {Herman}},
  \bibinfo {author} {\bibfnamefont {R.}~\bibnamefont {Shaydulin}}, \bibinfo
  {author} {\bibfnamefont {J.}~\bibnamefont {Basso}}, \bibinfo {author}
  {\bibfnamefont {S.}~\bibnamefont {Chakrabarti}}, \bibinfo {author}
  {\bibfnamefont {Y.}~\bibnamefont {Sun}},\ and\ \bibinfo {author}
  {\bibfnamefont {M.}~\bibnamefont {Pistoia}},\ }\bibfield  {title} {\bibinfo
  {title} {Parameter {S}etting in {Q}uantum {A}pproximate {O}ptimization of
  {W}eighted {P}roblems},\ }\href {https://doi.org/10.22331/q-2024-01-18-1231}
  {\bibfield  {journal} {\bibinfo  {journal} {{Quantum}}\ }\textbf {\bibinfo
  {volume} {8}},\ \bibinfo {pages} {1231} (\bibinfo {year} {2024})}\BibitemShut
  {NoStop}%
\bibitem [{\citenamefont {Yao}\ \emph {et~al.}(2020)\citenamefont {Yao},
  \citenamefont {Bukov},\ and\ \citenamefont {Lin}}]{yao2020policy}%
  \BibitemOpen
  \bibfield  {author} {\bibinfo {author} {\bibfnamefont {J.}~\bibnamefont
  {Yao}}, \bibinfo {author} {\bibfnamefont {M.}~\bibnamefont {Bukov}},\ and\
  \bibinfo {author} {\bibfnamefont {L.}~\bibnamefont {Lin}},\ }\href@noop {}
  {\bibinfo {title} {Policy gradient based quantum approximate optimization
  algorithm}} (\bibinfo {year} {2020}),\ \Eprint
  {https://arxiv.org/abs/2002.01068} {arXiv:2002.01068 [quant-ph]} \BibitemShut
  {NoStop}%
\bibitem [{\citenamefont {Zhu}\ \emph {et~al.}(2022)\citenamefont {Zhu},
  \citenamefont {Tang}, \citenamefont {Barron}, \citenamefont
  {Calderon-Vargas}, \citenamefont {Mayhall}, \citenamefont {Barnes},\ and\
  \citenamefont {Economou}}]{zhu2022adaptive}%
  \BibitemOpen
  \bibfield  {author} {\bibinfo {author} {\bibfnamefont {L.}~\bibnamefont
  {Zhu}}, \bibinfo {author} {\bibfnamefont {H.~L.}\ \bibnamefont {Tang}},
  \bibinfo {author} {\bibfnamefont {G.~S.}\ \bibnamefont {Barron}}, \bibinfo
  {author} {\bibfnamefont {F.~A.}\ \bibnamefont {Calderon-Vargas}}, \bibinfo
  {author} {\bibfnamefont {N.~J.}\ \bibnamefont {Mayhall}}, \bibinfo {author}
  {\bibfnamefont {E.}~\bibnamefont {Barnes}},\ and\ \bibinfo {author}
  {\bibfnamefont {S.~E.}\ \bibnamefont {Economou}},\ }\href@noop {} {\bibinfo
  {title} {An adaptive quantum approximate optimization algorithm for solving
  combinatorial problems on a quantum computer}} (\bibinfo {year} {2022}),\
  \Eprint {https://arxiv.org/abs/2005.10258} {arXiv:2005.10258 [quant-ph]}
  \BibitemShut {NoStop}%
\bibitem [{\citenamefont {{Harrigan \emph{et al.}}}(2021)}]{Harrigan_2021}%
  \BibitemOpen
  \bibfield  {author} {\bibinfo {author} {\bibfnamefont {M.~P.}\ \bibnamefont
  {{Harrigan \emph{et al.}}}},\ }\bibfield  {title} {\bibinfo {title} {Quantum
  approximate optimization of non-planar graph problems on a planar
  superconducting processor},\ }\href
  {https://doi.org/10.1038/s41567-020-01105-y} {\bibfield  {journal} {\bibinfo
  {journal} {Nature Physics}\ }\textbf {\bibinfo {volume} {17}},\ \bibinfo
  {pages} {332–336} (\bibinfo {year} {2021})}\BibitemShut {NoStop}%
\bibitem [{\citenamefont {{Weidenfeller}}\ \emph {et~al.}(2022)\citenamefont
  {{Weidenfeller}}, \citenamefont {{Valor}}, \citenamefont {{Gacon}},
  \citenamefont {{Tornow}}, \citenamefont {{Bello}}, \citenamefont
  {{Woerner}},\ and\ \citenamefont {{Egger}}}]{weidenfeller2022scaling}%
  \BibitemOpen
  \bibfield  {author} {\bibinfo {author} {\bibfnamefont {J.}~\bibnamefont
  {{Weidenfeller}}}, \bibinfo {author} {\bibfnamefont {L.~C.}\ \bibnamefont
  {{Valor}}}, \bibinfo {author} {\bibfnamefont {J.}~\bibnamefont {{Gacon}}},
  \bibinfo {author} {\bibfnamefont {C.}~\bibnamefont {{Tornow}}}, \bibinfo
  {author} {\bibfnamefont {L.}~\bibnamefont {{Bello}}}, \bibinfo {author}
  {\bibfnamefont {S.}~\bibnamefont {{Woerner}}},\ and\ \bibinfo {author}
  {\bibfnamefont {D.~J.}\ \bibnamefont {{Egger}}},\ }\bibfield  {title}
  {\bibinfo {title} {{Scaling of the quantum approximate optimization algorithm
  on superconducting qubit based hardware}},\ }\href@noop {} {\bibfield
  {journal} {\bibinfo  {journal} {arXiv e-prints}\ ,\ \bibinfo {eid}
  {arXiv:2202.03459}} (\bibinfo {year} {2022})},\ \Eprint
  {https://arxiv.org/abs/2202.03459} {arXiv:2202.03459 [quant-ph]} \BibitemShut
  {NoStop}%
\bibitem [{\citenamefont {Wurtz}\ \emph {et~al.}(2024)\citenamefont {Wurtz},
  \citenamefont {Sack},\ and\ \citenamefont {Wang}}]{wurtz2024solving}%
  \BibitemOpen
  \bibfield  {author} {\bibinfo {author} {\bibfnamefont {J.}~\bibnamefont
  {Wurtz}}, \bibinfo {author} {\bibfnamefont {S.}~\bibnamefont {Sack}},\ and\
  \bibinfo {author} {\bibfnamefont {S.-T.}\ \bibnamefont {Wang}},\ }\href@noop
  {} {\bibinfo {title} {Solving non-native combinatorial optimization problems
  using hybrid quantum-classical algorithms}} (\bibinfo {year} {2024}),\
  \Eprint {https://arxiv.org/abs/2403.03153} {arXiv:2403.03153 [quant-ph]}
  \BibitemShut {NoStop}%
\bibitem [{\citenamefont {Ebadi}\ \emph {et~al.}(2022)\citenamefont {Ebadi}
  \emph {et~al.}}]{Ebadi_2022}%
  \BibitemOpen
  \bibfield  {author} {\bibinfo {author} {\bibfnamefont {S.}~\bibnamefont
  {Ebadi}} \emph {et~al.},\ }\bibfield  {title} {\bibinfo {title} {{Quantum
  Optimization of Maximum Independent Set using Rydberg Atom Arrays}},\ }\href
  {https://doi.org/10.1126/science.abo6587} {\bibfield  {journal} {\bibinfo
  {journal} {Science}\ }\textbf {\bibinfo {volume} {376}},\ \bibinfo {pages}
  {1209} (\bibinfo {year} {2022})},\ \Eprint {https://arxiv.org/abs/2202.09372}
  {arXiv:2202.09372 [quant-ph]} \BibitemShut {NoStop}%
\bibitem [{\citenamefont {{Crooks}}(2018)}]{crooks2018performance}%
  \BibitemOpen
  \bibfield  {author} {\bibinfo {author} {\bibfnamefont {G.~E.}\ \bibnamefont
  {{Crooks}}},\ }\bibfield  {title} {\bibinfo {title} {{Performance of the
  Quantum Approximate Optimization Algorithm on the Maximum Cut Problem}},\
  }\href@noop {} {\bibfield  {journal} {\bibinfo  {journal} {arXiv e-prints}\
  ,\ \bibinfo {eid} {arXiv:1811.08419}} (\bibinfo {year} {2018})},\ \Eprint
  {https://arxiv.org/abs/1811.08419} {arXiv:1811.08419 [quant-ph]} \BibitemShut
  {NoStop}%
\bibitem [{\citenamefont {Zhou}\ \emph {et~al.}(2020)\citenamefont {Zhou},
  \citenamefont {Wang}, \citenamefont {Choi}, \citenamefont {Pichler},\ and\
  \citenamefont {Lukin}}]{zhou2018quantum}%
  \BibitemOpen
  \bibfield  {author} {\bibinfo {author} {\bibfnamefont {L.}~\bibnamefont
  {Zhou}}, \bibinfo {author} {\bibfnamefont {S.-T.}\ \bibnamefont {Wang}},
  \bibinfo {author} {\bibfnamefont {S.}~\bibnamefont {Choi}}, \bibinfo {author}
  {\bibfnamefont {H.}~\bibnamefont {Pichler}},\ and\ \bibinfo {author}
  {\bibfnamefont {M.~D.}\ \bibnamefont {Lukin}},\ }\bibfield  {title} {\bibinfo
  {title} {Quantum approximate optimization algorithm: Performance, mechanism,
  and implementation on near-term devices},\ }\href
  {https://doi.org/10.1103/PhysRevX.10.021067} {\bibfield  {journal} {\bibinfo
  {journal} {Phys. Rev. X}\ }\textbf {\bibinfo {volume} {10}},\ \bibinfo
  {pages} {021067} (\bibinfo {year} {2020})}\BibitemShut {NoStop}%
\bibitem [{\citenamefont {{Sack}}\ and\ \citenamefont
  {{Serbyn}}(2021)}]{sack2021quantum}%
  \BibitemOpen
  \bibfield  {author} {\bibinfo {author} {\bibfnamefont {S.~H.}\ \bibnamefont
  {{Sack}}}\ and\ \bibinfo {author} {\bibfnamefont {M.}~\bibnamefont
  {{Serbyn}}},\ }\bibfield  {title} {\bibinfo {title} {{Quantum annealing
  initialization of the quantum approximate optimization algorithm}},\
  }\href@noop {} {\bibfield  {journal} {\bibinfo  {journal} {arXiv e-prints}\
  ,\ \bibinfo {eid} {arXiv:2101.05742}} (\bibinfo {year} {2021})},\ \Eprint
  {https://arxiv.org/abs/2101.05742} {arXiv:2101.05742 [quant-ph]} \BibitemShut
  {NoStop}%
\bibitem [{\citenamefont {{Jain}}\ \emph {et~al.}(2021)\citenamefont {{Jain}},
  \citenamefont {{Coyle}}, \citenamefont {{Kashefi}},\ and\ \citenamefont
  {{Kumar}}}]{jain2021graph}%
  \BibitemOpen
  \bibfield  {author} {\bibinfo {author} {\bibfnamefont {N.}~\bibnamefont
  {{Jain}}}, \bibinfo {author} {\bibfnamefont {B.}~\bibnamefont {{Coyle}}},
  \bibinfo {author} {\bibfnamefont {E.}~\bibnamefont {{Kashefi}}},\ and\
  \bibinfo {author} {\bibfnamefont {N.}~\bibnamefont {{Kumar}}},\ }\bibfield
  {title} {\bibinfo {title} {{Graph neural network initialisation of quantum
  approximate optimisation}},\ }\href@noop {} {\bibfield  {journal} {\bibinfo
  {journal} {arXiv e-prints}\ ,\ \bibinfo {eid} {arXiv:2111.03016}} (\bibinfo
  {year} {2021})},\ \Eprint {https://arxiv.org/abs/2111.03016}
  {arXiv:2111.03016 [quant-ph]} \BibitemShut {NoStop}%
\bibitem [{\citenamefont {Sack}\ \emph {et~al.}(2023)\citenamefont {Sack},
  \citenamefont {Medina}, \citenamefont {Kueng},\ and\ \citenamefont
  {Serbyn}}]{Sack2022}%
  \BibitemOpen
  \bibfield  {author} {\bibinfo {author} {\bibfnamefont {S.~H.}\ \bibnamefont
  {Sack}}, \bibinfo {author} {\bibfnamefont {R.~A.}\ \bibnamefont {Medina}},
  \bibinfo {author} {\bibfnamefont {R.}~\bibnamefont {Kueng}},\ and\ \bibinfo
  {author} {\bibfnamefont {M.}~\bibnamefont {Serbyn}},\ }\bibfield  {title}
  {\bibinfo {title} {Recursive greedy initialization of the quantum approximate
  optimization algorithm with guaranteed improvement},\ }\href
  {https://doi.org/10.1103/PhysRevA.107.062404} {\bibfield  {journal} {\bibinfo
   {journal} {Phys. Rev. A}\ }\textbf {\bibinfo {volume} {107}},\ \bibinfo
  {pages} {062404} (\bibinfo {year} {2023})}\BibitemShut {NoStop}%
\bibitem [{\citenamefont {Day}\ \emph {et~al.}(2019)\citenamefont {Day},
  \citenamefont {Bukov}, \citenamefont {Weinberg}, \citenamefont {Mehta},\ and\
  \citenamefont {Sels}}]{bukov_2019}%
  \BibitemOpen
  \bibfield  {author} {\bibinfo {author} {\bibfnamefont {A.~G.~R.}\
  \bibnamefont {Day}}, \bibinfo {author} {\bibfnamefont {M.}~\bibnamefont
  {Bukov}}, \bibinfo {author} {\bibfnamefont {P.}~\bibnamefont {Weinberg}},
  \bibinfo {author} {\bibfnamefont {P.}~\bibnamefont {Mehta}},\ and\ \bibinfo
  {author} {\bibfnamefont {D.}~\bibnamefont {Sels}},\ }\bibfield  {title}
  {\bibinfo {title} {Glassy phase of optimal quantum control},\ }\href
  {https://doi.org/10.1103/PhysRevLett.122.020601} {\bibfield  {journal}
  {\bibinfo  {journal} {Phys. Rev. Lett.}\ }\textbf {\bibinfo {volume} {122}},\
  \bibinfo {pages} {020601} (\bibinfo {year} {2019})}\BibitemShut {NoStop}%
\bibitem [{\citenamefont {Broyden}(1970)}]{bfgs1}%
  \BibitemOpen
  \bibfield  {author} {\bibinfo {author} {\bibfnamefont {C.~G.}\ \bibnamefont
  {Broyden}},\ }\bibfield  {title} {\bibinfo {title} {{The Convergence of a
  Class of Double-rank Minimization Algorithms 1. General Considerations}},\
  }\href {https://doi.org/10.1093/imamat/6.1.76} {\bibfield  {journal}
  {\bibinfo  {journal} {IMA Journal of Applied Mathematics}\ }\textbf {\bibinfo
  {volume} {6}},\ \bibinfo {pages} {76} (\bibinfo {year} {1970})}\BibitemShut
  {NoStop}%
\bibitem [{\citenamefont {Fletcher}(1970)}]{bfgs2}%
  \BibitemOpen
  \bibfield  {author} {\bibinfo {author} {\bibfnamefont {R.}~\bibnamefont
  {Fletcher}},\ }\bibfield  {title} {\bibinfo {title} {{A new approach to
  variable metric algorithms}},\ }\href
  {https://doi.org/10.1093/comjnl/13.3.317} {\bibfield  {journal} {\bibinfo
  {journal} {The Computer Journal}\ }\textbf {\bibinfo {volume} {13}},\
  \bibinfo {pages} {317} (\bibinfo {year} {1970})}\BibitemShut {NoStop}%
\bibitem [{\citenamefont {Goldfarb}(1970)}]{bfgs3}%
  \BibitemOpen
  \bibfield  {author} {\bibinfo {author} {\bibfnamefont {D.}~\bibnamefont
  {Goldfarb}},\ }\bibfield  {title} {\bibinfo {title} {A family of
  variable-metric methods derived by variational means},\ }\href
  {http://www.jstor.org/stable/2004873} {\bibfield  {journal} {\bibinfo
  {journal} {Mathematics of Computation}\ }\textbf {\bibinfo {volume} {24}},\
  \bibinfo {pages} {23} (\bibinfo {year} {1970})}\BibitemShut {NoStop}%
\bibitem [{\citenamefont {Shanno}(1970)}]{bfgs4}%
  \BibitemOpen
  \bibfield  {author} {\bibinfo {author} {\bibfnamefont {D.~F.}\ \bibnamefont
  {Shanno}},\ }\bibfield  {title} {\bibinfo {title} {Conditioning of
  quasi-newton methods for function minimization},\ }\href
  {http://www.jstor.org/stable/2004840} {\bibfield  {journal} {\bibinfo
  {journal} {Mathematics of Computation}\ }\textbf {\bibinfo {volume} {24}},\
  \bibinfo {pages} {647} (\bibinfo {year} {1970})}\BibitemShut {NoStop}%
\bibitem [{\citenamefont {Bezanson}\ \emph {et~al.}(2017)\citenamefont
  {Bezanson}, \citenamefont {Edelman}, \citenamefont {Karpinski},\ and\
  \citenamefont {Shah}}]{julia}%
  \BibitemOpen
  \bibfield  {author} {\bibinfo {author} {\bibfnamefont {J.}~\bibnamefont
  {Bezanson}}, \bibinfo {author} {\bibfnamefont {A.}~\bibnamefont {Edelman}},
  \bibinfo {author} {\bibfnamefont {S.}~\bibnamefont {Karpinski}},\ and\
  \bibinfo {author} {\bibfnamefont {V.~B.}\ \bibnamefont {Shah}},\ }\bibfield
  {title} {\bibinfo {title} {Julia: A fresh approach to numerical computing},\
  }\href {https://doi.org/10.1137/141000671} {\bibfield  {journal} {\bibinfo
  {journal} {SIAM Review}\ }\textbf {\bibinfo {volume} {59}},\ \bibinfo {pages}
  {65} (\bibinfo {year} {2017})},\ \Eprint
  {https://arxiv.org/abs/https://doi.org/10.1137/141000671}
  {https://doi.org/10.1137/141000671} \BibitemShut {NoStop}%
\bibitem [{\citenamefont {Medina}(2024)}]{qaoa_landscapes}%
  \BibitemOpen
  \bibfield  {author} {\bibinfo {author} {\bibfnamefont {R.~A.}\ \bibnamefont
  {Medina}},\ }\href@noop {} {\bibinfo {title} {{QAOALandscapes.jl}}},\
  \bibinfo {howpublished}
  {\url{https://github.com/RaimelMedina/QAOALandscapes/tree/main}} (\bibinfo
  {year} {2024})\BibitemShut {NoStop}%
\bibitem [{\citenamefont {Besard}\ \emph {et~al.}(2018)\citenamefont {Besard},
  \citenamefont {Foket},\ and\ \citenamefont {De~Sutter}}]{besard2018juliagpu}%
  \BibitemOpen
  \bibfield  {author} {\bibinfo {author} {\bibfnamefont {T.}~\bibnamefont
  {Besard}}, \bibinfo {author} {\bibfnamefont {C.}~\bibnamefont {Foket}},\ and\
  \bibinfo {author} {\bibfnamefont {B.}~\bibnamefont {De~Sutter}},\ }\bibfield
  {title} {\bibinfo {title} {Effective extensible programming: Unleashing
  {Julia} on {GPUs}},\ }\bibfield  {journal} {\bibinfo  {journal} {IEEE
  Transactions on Parallel and Distributed Systems}\ }\href
  {https://doi.org/10.1109/TPDS.2018.2872064} {10.1109/TPDS.2018.2872064}
  (\bibinfo {year} {2018}),\ \Eprint {https://arxiv.org/abs/1712.03112}
  {arXiv:1712.03112 [cs.PL]} \BibitemShut {NoStop}%
\bibitem [{\citenamefont {Besard}\ \emph {et~al.}(2019)\citenamefont {Besard},
  \citenamefont {Churavy}, \citenamefont {Edelman},\ and\ \citenamefont
  {De~Sutter}}]{besard2019prototyping}%
  \BibitemOpen
  \bibfield  {author} {\bibinfo {author} {\bibfnamefont {T.}~\bibnamefont
  {Besard}}, \bibinfo {author} {\bibfnamefont {V.}~\bibnamefont {Churavy}},
  \bibinfo {author} {\bibfnamefont {A.}~\bibnamefont {Edelman}},\ and\ \bibinfo
  {author} {\bibfnamefont {B.}~\bibnamefont {De~Sutter}},\ }\bibfield  {title}
  {\bibinfo {title} {Rapid software prototyping for heterogeneous and
  distributed platforms},\ }\href@noop {} {\bibfield  {journal} {\bibinfo
  {journal} {Advances in Engineering Software}\ }\textbf {\bibinfo {volume}
  {132}},\ \bibinfo {pages} {29} (\bibinfo {year} {2019})}\BibitemShut
  {NoStop}%
\bibitem [{\citenamefont {Besard}\ and\ \citenamefont {Hawkins}()}]{metal_jl}%
  \BibitemOpen
  \bibfield  {author} {\bibinfo {author} {\bibfnamefont {T.}~\bibnamefont
  {Besard}}\ and\ \bibinfo {author} {\bibfnamefont {M.}~\bibnamefont
  {Hawkins}},\ }\href {https://doi.org/10.5281/zenodo.7139374} {\bibinfo
  {title} {Metal.jl}},\ \bibinfo {note} {available online at
  https://github.com/JuliaGPU/Metal.jl}\BibitemShut {NoStop}%
\bibitem [{\citenamefont {Mogensen}\ and\ \citenamefont
  {Riseth}(2018)}]{optim}%
  \BibitemOpen
  \bibfield  {author} {\bibinfo {author} {\bibfnamefont {P.~K.}\ \bibnamefont
  {Mogensen}}\ and\ \bibinfo {author} {\bibfnamefont {A.~N.}\ \bibnamefont
  {Riseth}},\ }\bibfield  {title} {\bibinfo {title} {Optim: A mathematical
  optimization package for julia},\ }\href
  {https://doi.org/10.21105/joss.00615} {\bibfield  {journal} {\bibinfo
  {journal} {Journal of Open Source Software}\ }\textbf {\bibinfo {volume}
  {3}},\ \bibinfo {pages} {615} (\bibinfo {year} {2018})}\BibitemShut {NoStop}%
\bibitem [{\citenamefont {Mogensen}\ \emph {et~al.}(2024)\citenamefont
  {Mogensen}, \citenamefont {White}, \citenamefont {Riseth}, \citenamefont
  {Holy}, \citenamefont {Lubin}, \citenamefont {Stocker}, \citenamefont
  {Noack}, \citenamefont {Levitt}, \citenamefont {Ortner}, \citenamefont
  {Legat}, \citenamefont {Johnson}, \citenamefont {Rackauckas}, \citenamefont
  {Yu}, \citenamefont {Carlsson}, \citenamefont {Lin}, \citenamefont
  {Strouwen}, \citenamefont {Grawitter}, \citenamefont {Arakaki}, \citenamefont
  {Pasquier}, \citenamefont {Covert}, \citenamefont {Rock}, \citenamefont
  {Creel}, \citenamefont {cossio}, \citenamefont {Regier}, \citenamefont
  {Kuhn}, \citenamefont {Stukalov}, \citenamefont {Williams},\ and\
  \citenamefont {Sato}}]{optim_2}%
  \BibitemOpen
  \bibfield  {author} {\bibinfo {author} {\bibfnamefont {P.~K.}\ \bibnamefont
  {Mogensen}}, \bibinfo {author} {\bibfnamefont {J.~M.}\ \bibnamefont {White}},
  \bibinfo {author} {\bibfnamefont {A.~N.}\ \bibnamefont {Riseth}}, \bibinfo
  {author} {\bibfnamefont {T.}~\bibnamefont {Holy}}, \bibinfo {author}
  {\bibfnamefont {M.}~\bibnamefont {Lubin}}, \bibinfo {author} {\bibfnamefont
  {C.}~\bibnamefont {Stocker}}, \bibinfo {author} {\bibfnamefont
  {A.}~\bibnamefont {Noack}}, \bibinfo {author} {\bibfnamefont
  {A.}~\bibnamefont {Levitt}}, \bibinfo {author} {\bibfnamefont
  {C.}~\bibnamefont {Ortner}}, \bibinfo {author} {\bibfnamefont
  {B.}~\bibnamefont {Legat}}, \bibinfo {author} {\bibfnamefont
  {B.}~\bibnamefont {Johnson}}, \bibinfo {author} {\bibfnamefont
  {C.}~\bibnamefont {Rackauckas}}, \bibinfo {author} {\bibfnamefont
  {Y.}~\bibnamefont {Yu}}, \bibinfo {author} {\bibfnamefont {K.}~\bibnamefont
  {Carlsson}}, \bibinfo {author} {\bibfnamefont {D.}~\bibnamefont {Lin}},
  \bibinfo {author} {\bibfnamefont {A.}~\bibnamefont {Strouwen}}, \bibinfo
  {author} {\bibfnamefont {J.}~\bibnamefont {Grawitter}}, \bibinfo {author}
  {\bibfnamefont {T.}~\bibnamefont {Arakaki}}, \bibinfo {author} {\bibfnamefont
  {B.}~\bibnamefont {Pasquier}}, \bibinfo {author} {\bibfnamefont {T.~R.}\
  \bibnamefont {Covert}}, \bibinfo {author} {\bibfnamefont {R.}~\bibnamefont
  {Rock}}, \bibinfo {author} {\bibfnamefont {M.}~\bibnamefont {Creel}},
  \bibinfo {author} {\bibnamefont {cossio}}, \bibinfo {author} {\bibfnamefont
  {J.}~\bibnamefont {Regier}}, \bibinfo {author} {\bibfnamefont
  {B.}~\bibnamefont {Kuhn}}, \bibinfo {author} {\bibfnamefont {A.}~\bibnamefont
  {Stukalov}}, \bibinfo {author} {\bibfnamefont {A.}~\bibnamefont {Williams}},\
  and\ \bibinfo {author} {\bibfnamefont {K.}~\bibnamefont {Sato}},\ }\href
  {https://doi.org/10.5281/zenodo.10858871} {\bibinfo {title}
  {Julianlsolvers/optim.jl: v1.9.3+doc1}} (\bibinfo {year} {2024})\BibitemShut
  {NoStop}%
\bibitem [{\citenamefont {Luo}\ \emph {et~al.}(2020)\citenamefont {Luo},
  \citenamefont {Liu}, \citenamefont {Zhang},\ and\ \citenamefont
  {Wang}}]{yao}%
  \BibitemOpen
  \bibfield  {author} {\bibinfo {author} {\bibfnamefont {X.-Z.}\ \bibnamefont
  {Luo}}, \bibinfo {author} {\bibfnamefont {J.-G.}\ \bibnamefont {Liu}},
  \bibinfo {author} {\bibfnamefont {P.}~\bibnamefont {Zhang}},\ and\ \bibinfo
  {author} {\bibfnamefont {L.}~\bibnamefont {Wang}},\ }\bibfield  {title}
  {\bibinfo {title} {Yao.jl: Extensible, efficient framework for quantum
  algorithm design},\ }\href {https://doi.org/10.22331/q-2020-10-11-341}
  {\bibfield  {journal} {\bibinfo  {journal} {Quantum}\ }\textbf {\bibinfo
  {volume} {4}},\ \bibinfo {pages} {341} (\bibinfo {year} {2020})}\BibitemShut
  {NoStop}%
\bibitem [{\citenamefont {Jones}\ and\ \citenamefont
  {Gacon}(2020)}]{jones_gradient}%
  \BibitemOpen
  \bibfield  {author} {\bibinfo {author} {\bibfnamefont {T.}~\bibnamefont
  {Jones}}\ and\ \bibinfo {author} {\bibfnamefont {J.}~\bibnamefont {Gacon}},\
  }\href@noop {} {\bibinfo {title} {Efficient calculation of gradients in
  classical simulations of variational quantum algorithms}} (\bibinfo {year}
  {2020}),\ \Eprint {https://arxiv.org/abs/2009.02823} {arXiv:2009.02823
  [quant-ph]} \BibitemShut {NoStop}%
\bibitem [{\citenamefont {Ostrowski}(1959)}]{ostrowski}%
  \BibitemOpen
  \bibfield  {author} {\bibinfo {author} {\bibfnamefont {A.~M.}\ \bibnamefont
  {Ostrowski}},\ }\bibfield  {title} {\bibinfo {title} {A quantitative
  formulation of sylvester's law of inertia},\ }\href
  {https://doi.org/10.1073/pnas.45.5.740} {\bibfield  {journal} {\bibinfo
  {journal} {Proceedings of the National Academy of Sciences}\ }\textbf
  {\bibinfo {volume} {45}},\ \bibinfo {pages} {740} (\bibinfo {year} {1959})},\
  \Eprint
  {https://arxiv.org/abs/https://www.pnas.org/doi/pdf/10.1073/pnas.45.5.740}
  {https://www.pnas.org/doi/pdf/10.1073/pnas.45.5.740} \BibitemShut {NoStop}%
\end{thebibliography}%
\end{document}